\documentclass[twocolumn,showpacs,prb,footinbib,a4paper,superscriptaddress,floatfix]{revtex4}

\usepackage{amssymb,amsmath}
\usepackage{graphics}
\usepackage{dcolumn}
\usepackage{bm}
\usepackage{epsfig}

\usepackage{feynmf}

\newcommand{\expect}[1]{\langle #1 \rangle\ }
\newcommand{\x}[1]{\underline{#1}}

\begin{document}

\hyphenation{meso-scopic nano-scale tech-nique}

\title{
$GW$ approximations and vertex corrections on the Keldysh time-loop
contour: application for model systems at equilibrium 
}

\author{H. Ness}
\email{herve.ness@york.ac.uk}
\affiliation{Department of Physics, University of York, Heslington, York YO10 5DD,
UK}
\affiliation{European Theoretical Spectroscopy Facility (ETSF)}

\author{L. K. Dash}
\affiliation{Department of Physics, University of York, Heslington, York YO10 5DD,
UK}
\affiliation{European Theoretical Spectroscopy Facility (ETSF)}

\author{M. Stankovski} \affiliation{IMCN-NAPS, Universit\'e Catholique
  de Louvain, Place Croix du Sud 1 bte 3, B-1348 Louvain-la-Neuve,
  Belgium} \affiliation{European Theoretical Spectroscopy Facility
  (ETSF)}

\author{R. W. Godby}
\affiliation{Department of Physics, University of York, Heslington, York YO10 5DD,
UK}
\affiliation{European Theoretical Spectroscopy Facility (ETSF)}

\date{\today}

\begin{abstract}
  We study the effects of self-consistency and vertex corrections on
  different $GW$-based approximations for model systems of interacting
  electrons.
  For dealing with the most general case, we use the Keldysh time-loop
  contour formalism to evaluate the single-particle Green's functions.
  We provide the formal extension of Hedin's $GW$ equations for
  the Green's function in the Keldysh formalism.
  We show an application of our
  formalism to the plasmon model of a core electron within the
  plasmon-pole approximation. We study in detail the effects of the diagrammatic
  perturbation expansion of the core-electron/plasmon coupling on the
  spectral functions in the so-called S-model. The S-model provides an exact
  solution at equilibrium for comparison with the diagrammatic expansion
  of the interaction.
  We show that self-consistency is essential in $GW$-based
  calculations to obtain the full spectral information. The
  second-order exchange diagram (i.e.\ a vertex correction)
  is also crucial to obtain the good spectral description of the plasmon
  satellites.
  We corroborate these results by considering conventional
  equilibrium $GW$-based calculations for the pure jellium model. We find
  that with no second-order vertex correction, one cannot obtain
  the full set of plasmon side-band resonances.  We also discuss in detail
  the formal expression of the Dyson equations obtained for the
  time-ordered Green's function at zero and finite temperature from
  the Keldysh formalism and from conventional equilibrium many-body perturbation theory.
\end{abstract}
 

\maketitle

\section{Introduction}
\label{sec:intro}

Equilibrium, zero- and finite-temperature Green's functions techniques
based on many-body perturbation theory (MBPT) are widely used in
electronic-structure and total energy calculations
\cite{Abrikosov:1963}.  Hedin's formulation
\cite{Hedin:1965,Hedin:1969} for the electronic Green's function
closes the many-body hierarchy by expanding the electron self-energy
of the one-particle Green's function in terms of the screened Coulomb
interaction in the presence of vertex corrections. 

Without these vertex
corrections, one obtains the conventional $GW$ equations
\cite{Hedin:1969,Godby:1988,DelSole:1994,Aryasetiawan:1998,
  Rieger:1999,Onida:2002}.
The $GW$ method is an approximate treatment of the propagation of
electrons: it can be seen as if electrons interact with themselves via
a Coulomb interaction that is screened by virtual electron-hole
pairs. In bulk semiconductors, the $GW$ approximation is known to lead
to surprisingly accurate band gaps
\cite{Godby:1988,Rohlfing:1995,Aryasetiawan:1998, Rieger:1999}, while
for finite-size systems and molecules the method provides
qualitatively correct values of ionization energies and electron
affinities \cite{Rohlfing:2000}.  It also provides a convenient
starting point for many useful approximations and applications to
photoemission spectroscopy \cite{Onida:2002} and optical absorption in
metals or semiconductors as well as in finite size molecular systems
\cite{Aryasetiawan:1998,Rohlfing:2000,Benedict:2002,Blase:2011,Faber:2011}.
Most practical $GW$ calculations today are performed in a perturbative
manner using equilibrium MBPT.

However, if we want to consider a system driven out of equilibrium by an 
external ``force'', such as, for
example, a molecular wire coupled to electrodes sustaining an electronic
current flow, or any system driven by an external electromagnetic field
(time-dependent or not), we need to extend the equations for the
dynamics of the quantum many-body interacting system (Hedin's
equations or their simplified $GW$ form) to non-equilibrium conditions.

For this, the non-equilibrium Green's function (NEGF)
technique\cite{Keldysh:1965,Wagner:1991,Danielewicz:1984,vanLeeuwen:2006}
has been widely used to calculate electronic transport properties of
mesoscopic\cite{Haug:1996} and
nanoscale\cite{Myohanen:2009,Myohanen:2008,Rammer:2007,Stefanucci:2004a,Rammer:1991,Rammer:1986}
systems, plasmas, quantum transport in semiconductors\cite{Haug:1996}
and high-energy processes in nuclear physics \cite{Danielewicz:1984b}.
Also known as the closed time-path formalism
\cite{Schwinger:1961,Chou:1985}, the NEGF formalism depends on an ``artificial'' time
parameter that runs on a mathematically convenient time-loop contour
(plus eventually an imaginary time for taking into account the initial
correlation and statistical boundary conditions).  It is a formal
procedure that only has a direct physical meaning when one projects
back the time parameters of the time-loop contour onto real times.  It
was introduced because it allows one to obtain self-consistent
Dyson-like equations for the Keldysh Green's function using
Schwinger's functional derivative technique.  Transforming the Dyson
equation to real time by varying the Keldysh time parameter over the
time-loop contour results in a set of self-consistent equations
for the different non-equilibrium Green's functions (advanced/retarded
or lesser/greater).
The NEGF technique is general and can treat non-equilibrium as well as
equilibrium conditions, and the zero and finite temperature limits, within
a single framework.

The NEGF technique has been applied to the study of different
levels of self-consistency in the $GW$ approach for atoms,
molecules and semiconductors in
Refs.~[\onlinecite{Spataru:2004,Stan:2006,Dahlen:2007,Stan:2009a,Stan:2009b,Thygesen:2007,Rostgaard:2010}].
However, these works did not include the effects of simultaneous
self-consistency and vertex corrections.
{ Other levels of approximation for electron-electron interactions have also been
considered in finite-size nanoclusters by using the Kadanoff-Baym
flavour of NEGF \cite{PuigvonFriesen:2009,PuigvonFriesen:2010}.}

In this paper, we want to study these effects (self-consistency and vertex
corrections) and use the most general formalism to deal with the full equivalent to Hedin's 
$GW$ equations. { We believe that the Keldysh formalism, even applied to equilibrium 
conditions, can be more useful than the conventional approaches since 
it is by nature a more general approach.}

We extend Hedin's equations to the Keldysh time-loop contour, and derive 
the equations for the one-particle Green's function $G$, self-energy $\Sigma$, 
screened Coulomb interaction $W$ and for the (3-point) vertex functions $\Gamma$.
Note that a non-equilibrium approach to Hedin's $GW$ equations has been provided
in Ref.~[\onlinecite{Harbola:2006}] where an alternative distinct approach 
based on the Liouvillian superoperator formalism is used.  
However, working in a Louivillian vector space is much less convenient and much more
computationaly demanding for practical applications than working within an Hilbert 
space as in the formalism we develop below.

We then apply our formalism to the calculation of the spectral function of a particular 
model of an homogeneous electron gas: the plasmon model for a core electron
\cite{Langreth:1970,Minnhagen:1975,Hedin:1969}.  We choose this model
as it can be solved exactly at equilibrium, 
and thus we are able to
compare the different approximations introduced in the calculations
(self-consistency versus one-shot calculations, and/or vertex corrections) and
check their validity for different limiting cases (the high and low electronic
density regimes).
We also compare the outcome of these calculations with conventional $GW$ calculations 
for the jellium model. We examine if the effects on the spectral functions
rendered by self-consistency iterations and the inclusion of vertex corrections
which we find for the plasmon model with a core electron also hold for the jellium model.

{  To our knowledge, the only available exact results are for
  equilibrium conditions, and thus we benchmark our formalism against
  exact results at equilibrium before extending the discussion to
  non-equilibrium conditions.}

The paper is organized as follows.
In Section \ref{sec:equiGW}, we recall the expressions of Hedin's 
$GW$ equations and briefly review the performance of conventional
$GW$ calculations.
The extension of Hedin's equations to the Keldysh time-loop
contour is provided in Section \ref{sec:NEGW}.
We also show that we recover the conventional
non-equilibrium $GW$ formalism developed and used by others
\cite{Spataru:2004,Stan:2009a,Stan:2009b,Thygesen:2007,Rostgaard:2010} when
ignoring the vertex corrections in Appendix \ref{app:lowestorder_exp}).
The lowest-order expansion, in terms of the interaction for the
screened Coulomb interaction $W$ and for the vertex functions $\Gamma$,
is given in Appendix \ref{app:lowestorder_exp}. In Appendix \ref{app:timeorderedGFs}
we also provide a rigorous mathematical proof of the difference between 
equilibrium time-ordered Green's functions in the zero and finite temperature
limits that were discussed less rigorously in Chapter IV.17 of Ref.~[\onlinecite{Hedin:1969}].

In Section \ref{sec:applic}, we apply our formalism
to the calculation of the spectral function of a model system and core electron
coupled to a plasmon mode
\cite{Langreth:1970,Minnhagen:1975,Hedin:1969}.  The exact solution of
this model at equilibrium permits us to examine the effects of self-consistency
and vertex corrections on the spectral density. We also examine these effects
for another model of an electron gas, the jellium model, by using 
conventional $GW$ calculations (Section \ref{martin-jellium}). 
We show a general trend: second-order diagrams for the interactions (i.e.~vertex
corrections) are necessary to obtain the full series of plasmon side-band peaks.
Finally, we conclude our work in Section \ref{ccl}.

\section{Hedin's GW equations}
\label{sec:equiGW}

Hedin's $GW$ equations \cite{Hedin:1965,Hedin:1969} were originally
derived for the time-ordered single-particle Green's function
$G$ at equilibrium, defined by
\begin{equation}
\label{eq:G++}
G(12)=-{\rm i}\expect{\mathcal{T} \Psi(1) \Psi^\dag(2)}.
\end{equation}
They are expressed as follows \cite{Hedin:1965,Hedin:1969}:
\begin{subequations}
\label{eq:HedinGW}
\begin{align}
G(12) & = G_0(12)+
\int {\rm d}(34)\ G_0(13)\ \Sigma(34)\ G(42), \label{eq:HedinGW-G} \\
\Sigma(12) & = {\rm i} \int {\rm d}(34)\ 
G(13)\ \Gamma(32;4)\ W(41), \label{eq:HedinG-Sig} \\
W(12) & = v(12)+
\int {\rm d}(34)\ v(13)\ \tilde{P}(34)\ W(42), \label{eq:HedinGW-W}\\
\tilde{P}(12) & = -{\rm i} \int {\rm d}(34)\
G(13)\  G(41)\ \Gamma(34;2), \label{eq:HedinGW-P}\\
\Gamma(12;3) & = \delta(12) \delta(13) + \label{eq:HedinGW-Gamma}\\
 & \qquad \int {\rm d}(4567)\ 
\frac{\delta \Sigma(12)}{\delta G(45)}\
G(46)\  G(75)\ \Gamma(67;3), \nonumber
\end{align}
\end{subequations}
with the usual notation for the space-time coordinates: any integer
$i$ represents a point in space-time
$\x{i}=\mathbf{x}_i=(\mathbf{r}_i,t_i)$; and for the electron
single-particle Green's function $G$, the corresponding self-energy
$\Sigma$, the screened Coulomb interaction $W$, the irreducible
polarizability $\tilde{P}$ (sometimes also called polarization), and
the vertex function $\Gamma$.

Up to now, most practical $GW$ calculations are performed not fully self-consistently,
using a single iteration of the $GW$ equations, called one-shot
$GW$ or $G_{0}W_{0}$. When a single iteration is performed, the
initial approximation must be good, so typically $G_{0}$ is
constructed from the orbitals of any current ground-state method which
correctly predicts the basic physics of the system.

The application of $G_{0}W_{0}$ corrections to
spectral properties and  band gaps as calculated in the local density
 approximation (LDA) and generalized gradient approximation (GGA) 
in density functional theory (DFT) is a long-standing 
success story~\cite{Aulbur2000}, at least for many \emph{s-p} bonded systems.
However, total energies calculated from the Galitskii-Migdal
formula at the $G_{0}W_{0}$ level are generally worse than those given
by other ground-state methods.
The initial close agreements with measured band gaps were
later shown to be partly due to technical approximations used along
the way. Several studies have shown that LDA+$G_{0}W_{0}$
systematically underestimates band gaps when solved in
state-of-the-art all-electron schemes with full explicit treatment of
frequency integrals (gaps are underestimated by about 1-10\% for
\emph{s-p} bonded systems and about 20-50\% for systems with \emph{d}
electrons, like rare-earth oxides, sulfides and
nitrides)\cite{Rinke2008,Schilfgaarde2006}. The remaining discrepancy
has prompted the search for more accurate but still tractable methods.

In general, attempts at fully self-consistent $GW$ have shown that
spectral properties worsen as compared with $G_{0}W_{0}$, while total
energies improve. $GW$ band gaps were first shown to be larger than
expected in a quasi one-dimensional Si wire model by de Groot \emph{et
  al}\cite{deGroot1995}.  Von Barth and Holm showed that in jellium for
a self-consistent update only of the Green's function (a $GW_{0}$
approach) a displacement of weight from quasiparticle peaks into the
incoherent background occurs~\cite{vonBarth1996}. Also, the occupied
bandwidth broadens rather than narrows, as expected from experiments
on simple metals. At the same time, E.~L. Shirley showed that the
bandwidth of jellium broadens further with full self-consistency,
while the bandwidth narrows again once vertex corrections are taken
into account~\cite{Shirley:1996}. The effects of non-locality in vertex
corrections were also addressed in Ref.[\onlinecite{Romaniello:2009}].

Later studies have shown that $GW$ total energies for jellium are very
accurate\cite{Holm1998,Holm1999,Pablo2001}, as expected from a
conserving approximation in the Baym-Kadanoff
sense~\cite{Baym1961}. This holds true even for low-dimensional atomic
and molecular systems, and ionization potentials as calculated by the
extended Koopman's theorem also tend to be
accurate~\cite{Stan:2006,Kaasbjerg2010,Rostgaard:2010}.  Full $GW$
calculations were performed by Kutepov \emph{et al.}\ for simple metals
and semiconductors, They showed {\it inter alia} that the calculated
equilibrium lattice parameters were all very close to the experimental
ones\cite{Kutepov2009}.

In view of improving the starting point, quasi-particle
self-consistent GW has emerged as a good
compromise between self-consistency and a practical path to good
spectral properties~\cite{Faleev2004}. It has been shown that vertex
corrections further improve the correspondence between theory and
experiment~\cite{Shishkin2007}, but consistently accurate results still
remain elusive for systems with localized states, defects and band
offsets\cite{Hong2009,Athanasios2007,Giantomassi2011}. 

In particular
relevance to the present paper, no existant implementation of $GW$
seems to describe the full spectrum of plasmon satellites in metals.

\section{Extension of Hedin's $GW$ equations to the Keldysh time-loop contour}
\label{sec:NEGW}

We now consider the generalization of the single-particle Green's function on
the time-loop contour (the so-called Keldysh contour $C_K$ with two
branches, branch ($+$) for forward time evolution and branch ($-$) for
backward time evolution):
\begin{equation}
G(12)=-{\rm i}\expect{\mathcal{T}_{C_K} \Psi(1) \Psi^\dag(2)} .
\end{equation}

For the moment, we do not specify the nature of the ``external force''
that drives the system out of equilibrium.  We consider the
generalized Green's function on the Keldysh time-loop contour and
hence end up with four different Keldysh components for the Green's
functions: $G^{++},G^{+-},G^{-+},G^{--}$, defined according to the way
the two real-time arguments $(t_1,t_2)$ are positioned on the
time-loop contour $C_K$.  The initial correlations (i.e\ the initial
boundary conditions) are assumed to be dealt with in an appropriate
way \cite{Wagner:1991,Stefanucci:2004a,vanLeeuwen:2006}.

To derive the NE-$GW$ equations, we proceed as follows: in each
integral $\int {\rm d}(1)$, the time is integrated over the time-loop
contour $C_K$: $\int_{C_K} {\rm d}\tau_1$ and then decomposed onto the
two real-time branches: $\int_{C_K} {\rm d}\tau_1 \equiv \int_{(+)}
{\rm d} t^+_1+\int_{(-)} {\rm d} t^-_1 = \int {\rm d} t^+_1 - \int
{\rm d} t^-_1$.  We then calculate the different components $X^{\eta_1
  \eta_2}$ (with $\eta_{1,2}=\pm$) for the Green's function,
self-energy, screened Coulomb interaction $W$, polarizability $P$,
and vertex function $\Gamma$.  Where possible, we re-express these in a
more convenient way by using the relations between the different Green's
functions and self-energies on the time-loop contour (see Appendix
\ref{app:GFSEdefinitions}).

There are actually three kinds of equation in Hedin's
$GW$ equations Eq.~(2a-e).  First, there is a set of Dyson-like equations
for the electron Green's function $G$ and for the boson Green's
function $W$, i.e.\ the screened Coulomb interaction. In these two
equations, the vertex function $\Gamma$ does not appear explicitly.
Next there is another set of equations for the electron self-energy
$\Sigma$ and for the polarizability (the boson self-energy)
$\tilde{P}$. In these equations, the vertex function appears
explicitly.  Finally there is the equation for the vertex function itself,
$\Gamma$. The vertex function can be expanded as a series
$\Gamma(12;3)=\sum_n \Gamma_{(n)}(12;3)$ where the index $n$
represents the number of times the screened Coulomb interaction $W$
appears explicitly in the series expansion.  Each occurrence of the 
screened Coulomb interaction $W$ in the vertex function $\Gamma$ is 
generated by the functional derivative $\delta\Sigma / \delta G$.

Finally, one should note that the equilibrium properties of the system
are, in principle, recovered from the extension of Hedin's $GW$
equations to the Keldysh time-loop contour when the external driving
force is omitted and the whole system is at thermodynamical
equilibrium.

\subsection{The electron Green's function and the self-energy}
\label{sec:GF+Sigma}

Following the prescriptions given above, we calculate the components
$G^{++}$, $G^{+-}$ and $G^{-+}$ from the extension of
Eq.~(\ref{eq:HedinGW}) on the time-loop contour, and we find the
Dyson-like equation for $G^{r,a}$:
\begin{equation}
G^{r,a}(12)=G_0^{r,a}(12)+
\int {\rm d}(34)\ G_0^{r,a}(13) \Sigma^{r,a}(34) G^{r,a}(42),
\end{equation}
which has the same functional form as in Eq.~(\ref{eq:HedinGW}).

We also obtain the following quantum kinetic equation (QKE) for $G^\lessgtr$:
\begin{equation}
\begin{split}
& G^\lessgtr(12)
=  \int {\rm d}(3456) \\
  & \left[
\delta(14)+G^r(13) \Sigma^r(34)
\right]
G_0^\lessgtr(45)
\left[
\delta(52)+\Sigma^a(56) G^a(62)
\right] \\
 & +
\int {\rm d}(34)\
G^r(13)\ \Sigma^\lessgtr(34)\ G^a(42).
\end{split}
\end{equation}

\subsection{The screened Coulomb potential}
\label{sec:W}

By looking at Eq.(2c), one can see that $W$ has the same
functional form as the electron Green's function $G$.  The
screened Coulomb interaction $W$ is a bosonic Green's function with an
associated bosonic self-energy, the polarizability $\tilde{P}$.  With
the formal equivalence $(G,\Sigma)\leftrightarrow (W,\tilde{P})$, one
can expect to obtain a Dyson-like equation for the advanced and
retarded screened Coulomb interaction and a quantum kinetic equation
for $W^\lessgtr$ as equivalently obtained for the electron Green's
function.

This is indeed what we find: $W^{r,a}$ follows the usual Dyson-like
equation as
\begin{equation}
W^{r,a}(12)=v(12)+\int {\rm d}(34)\ v(13)\ \tilde{P}^{r,a}(34)\ W^{r,a}(42),
\end{equation}
or in a more compact notation 
\begin{equation}
\begin{split}
W^{r,a} & = 
 v+v\tilde{P}^{r,a} W^{r,a}=v+W^{r,a}\tilde{P}^{r,a}v \\
&= 
 v[1-\tilde{P}^{r,a}v]^{-1}=[1-v\tilde{P}^{r,a}]^{-1}v,
\end{split}
\end{equation}
where any product $XY$ implies
a space-time integration $[XY](12)=\int {\rm d}(3) X(13)Y(32)$.

Since the bare Coulomb potential $v(12)$ is instantaneous, it corresponds to
an interaction local in time and therefore its extension to the Keldysh contour
has no $v^{+-}$ or $v^{-+}$ components. Hence, we obtain the following quantum
kinetic equations for $W^\lessgtr$ :
\begin{equation}
W^\lessgtr(12)= \int {\rm d}(34)\ W^r(13)\ \tilde{P}^\lessgtr(34)\ W^a(42) .
\end{equation}

\subsection{The vertex function $\Gamma(12;3)$ on the contour $C_K$}
\label{subsec:fullGamma}

The derivation of $\Gamma(12;3)$ on $C_K$ does not create any formal
difficulties.  However since $\Gamma(12;3)$ is a three-point function,
it is not possible to recover a Dyson-like or a quantum-kinetic-like
equation for $\Gamma$.

For any Keldysh components of the vertex function $\Gamma^{\eta_3
  \eta_2 \eta_4}(32;4)$, we can formally write the different
components of the self-energy on the Keldysh contour as follows:
\begin{equation}
\begin{split}
\Sigma^{\eta_1 \eta_2}(12)={\rm i} \sum_{\eta_3 \eta_4} \ & \eta_3 \eta_4
\int {\rm d}(34) \\
 & G^{\eta_1 \eta_3}(13)\ \Gamma^{\eta_3 \eta_2 \eta_4}(32;4)\ W^{\eta_4 \eta_1}(41) ,
\end{split}
\end{equation}
and likewise for the polarizability
\begin{equation}
\begin{split}
\tilde{P}^{\eta_1 \eta_2}(12)=-{\rm i} \sum_{\eta_3 \eta_4} \  & \eta_3 \eta_4
\int {\rm d}(34) \\
& G^{\eta_1 \eta_3}(13)\ G^{\eta_4 \eta_1}(41)\ \Gamma^{\eta_3 \eta_4 \eta_2}(34;2) .
\end{split}\end{equation}

Now we need to close the above equations, i.e.\ to find an equation for
the different components $\Gamma^{\eta_1 \eta_2 \eta_3}(12;3)$ of the
vertex function.  By considering the equivalent of
Eq.(2e) on the Keldysh contour, we obtain
\begin{equation}
\begin{split}
& \Gamma^{\eta_1 \eta_2 \eta_3}(12;3)=
 \delta^{\eta_1 \eta_2}(12) \delta^{\eta_1 \eta_3}(13)
+
\sum_{\eta_4 ... \eta_7} \eta_4 \eta_5 \eta_6 \eta_7 \\
& \int {\rm d}(4567)\ 
\frac{\delta \Sigma^{\eta_1 \eta_2}(12)}{\delta G^{\eta_4 \eta_5}(45)}
G^{\eta_4 \eta_6}(46)  G^{\eta_7 \eta_5}(75) 
\Gamma^{\eta_6 \eta_7 \eta_3}(67;3) .
\end{split}
\end{equation}

In Appendix \ref{app:lowestorder_exp}, we consider the series expansion 
of the vertex function 
$\Gamma(12;3)=\sum_n \Gamma_{(n)}(12;3)$ where the index $n$
represents the number of times the screened Coulomb interaction $W$
appears explicitly in the series expansion, and we provide explicit
results for the electron self-energy $\Sigma$ and polarizability $P$ for the 
lowest order terms  $\Gamma_{(0)}(12;3)$ and $\Gamma_{(1)}(12;3)$.

\section{Application to models related to the homogeneous electron gas}
\label{sec:applic}

Now we want to test our extended formalism of Hedin's $GW$ equation
onto the Keldysh time-loop contour and the corresponding series
expansion of the vertex functions.  The importance of self-consistency
and vertex corrections was discussed in Section \ref{sec:equiGW}.
Self-consistency and vertex corrections apply in both equilibrium and
non-equilibrium systems and therefore are more conveniently addressed
in as simple a model system as possible.

Calculations could be performed for several model systems, but would
not lead to any pertinent conclusions if they could not be compared to
exact results.
To our knowledge, exact results for interacting electron systems are few and
not as widespread as numerical (highly accurate) calculations even for
models of interacting electron systems.
One of the available exactly-solvable models has been used in the context of 
x-ray spectroscopy of metals, and leads to tractable analytical expressions
for the electron Green's function: the plasmon model for the core electron
\cite{Langreth:1970}.

In the next section, we consider this exactly-solvable model and compare the
exact results with those obtained from our $GW$ formalism, at zero and
finite temperatures and with
or without lowest-order vertex corrections.
We note here that the exact solution is obtained for a model of an
homogeneous electron gas \emph{at equilibrium}. Dealing with an
interacting system at equilibrium does not cause any problem within our
formalism, since the equilibrium condition is just a special case of
our more general formalism for non-equilibrium conditions (See
appendix \ref{app:timeorderedGFs} for a full discussion about the 
equilibrium limit of the Keldysh formalism at zero and
finite temperatures).

\subsection{Effective Hamiltonian for the plasmon model of a core electron}
\label{sec:Heff}

The properties of an homogeneous 3D electron gas can be well-described
within the plasmon model. The plasmon model is defined from Hedin's
equations Eqs.~(2a-e) together with the so-called
plasmon-pole parametrization.  In reciprocal space, the screened
Coulomb potential can be written as $W(\omega,q)=v_q\
\epsilon^{-1}(\omega,q)$, where $v_q$ is the Fourier component $q$ of
the Coulomb potential. The dielectric function $\epsilon^{-1}(\omega,q)$ is then obtained from the plasmon-pole
approximation $\epsilon^{-1}(\omega,q)=1+\omega_p^2/(\omega^2-\omega_q^2)$, where
$\omega_p$ is the bulk plasmon energy, related to the electron density
$n$ as usual, $\omega_p^2 = (4\pi n e^2 / m)$, and the plasmon
dispersion $\omega_q$ remains to be defined.

Within this model, the dynamic part of the Coulomb potential $W(\omega,q)-v_q$ 
can be re-expressed as
\begin{equation}
v_2 = v_q \left( \epsilon^{-1}(\omega,q) -1 \right)
= \frac{v_q \omega_p^2}{2\omega_q}\ \frac{2\omega_q}{\omega^2-\omega_q^2}
= \gamma_q^2\ B(\omega,q),
\label{eq:boson_propagator}
\end{equation}
which involves a coupling constant $\gamma_q$ and the bosonic propagator
$B(\omega,q)$ of the plasmon modes.

Following Refs.~[\onlinecite{Minnhagen:1975,Langreth:1970,Hedin:1969}]
we consider the following Hamiltonian for the plasmon model of a core
electron
\begin{equation}
H_{\rm eff} = \varepsilon_c c^\dag c + \sum_q \omega_q b_q^\dag b_q 
            + \sum_q  \gamma_q c^\dag c (b_q + b_{-q}^\dag).  
\end{equation}
For this model of the core-electron case there exists a precise and
well-defined relation between the solution defined by a plasmon model
for an electron gas and the solution defined by the corresponding
effective Hamiltonian $H_{\rm eff}$\cite{Minnhagen:1975}.
Finally we consider the $q \rightarrow 0$ limit of static
random-phase approximation \cite{Hedin:1969} for the plasmon dispersion:
\begin{equation}
\omega_q  = \omega_p \left( \frac{q^4}{(\omega_p^0)^2} 
		      	  + \frac{16}{3} \frac{q^2}{(\omega_p^0)^2} 
			  + 1 \right)^{1/2},  
\end{equation}
with $\omega_p^0=\omega_p/\varepsilon_F=4(\frac{\alpha
  r_S}{3\pi})^{1/2}$, $\alpha=(\frac{4}{9\pi})^{1/3}$, and $r_S$
defines the electron density $n=(\frac{4\pi}{3}r_S^3)^{-1}$.

\subsection{The S-model}
\label{sec:Smodel}

A particularly simple model of a core electron, known as the S-model \cite{Minnhagen:1975},
is obtained by further replacing $\omega_q^{-1}$ by a step function
$\omega_q^{-1} \rightarrow \omega_p^{-1} \theta(q_c-q)$, where the
cut-off parameter $q_c$ is determined by:
\begin{equation}
q_c = \int_0^{q_c} dq = \int_0^\infty \frac{\omega_p^2}{\omega_q^2}\ dq .
\end{equation}
From this definition of $q_c$, it follows that the energy shift parameter 
\begin{equation}
D=\sum_q \frac{\gamma_q^2}{\omega_q} = \frac{1}{2} \sum_q v_q \frac{\omega_p^2}{\omega_q^2} ,
\end{equation}
is the same as for the corresponding plasmon model.

The solution of the S-model can be mapped onto a simpler
Hamiltonian, giving rise to the same spectral information
\begin{equation}
H_{\rm eff} = \varepsilon_c c^\dag c + \omega_p  b^\dag b 
            + \gamma_0 c^\dag c (b + b^\dag),  
\label{eq:Smodel}
\end{equation}
with $\gamma_0^2 = D \omega_p$.

An analytical expression for the relaxation energy $D$ is found
from the chosen dispersion relation of the plasmon frequency
$\omega_q$.

We then find that the corresponding relaxation energy is given by
\begin{equation}
D = \sum_q \frac{\gamma_q^2}{\omega_q} 
= \frac{1}{2} \int \frac{{\rm d}^3 q}{(2\pi)^3} v_q \frac{\omega_p^2}{\omega_q^2}
= \frac{1}{2\sqrt{2}}\frac{\omega_p^0}{(\omega_p^0+\frac{8}{3})^{1/2}} .
\end{equation}

This result is very similar to the relaxation energy found by
Minnhagen \cite{Minnhagen:1975} when one replaces the prefactor
$\frac{16}{3}$ in the dispersion relation $\omega_q$ by $\frac{4}{3}$
and when one uses the trigonometric relations
$\sin(\frac{a}{2})=\sqrt{\frac{1-\cos a}{2}}$ and
$\cos[\tan^{-1}(u)]=1/\sqrt{(1+u^2)}$.

The other advantage of dealing with the S-model is that it has an
exact solution \cite{Langreth:1970,Minnhagen:1975,Ness:2006} which can
be compared with approximate calculations performed with Hedin's $GW$
equation for different levels of expansion of the self-energy and/or
vertex function.
The exact solution of the S-model at zero temperature provides us with
an analytical expression for the retarded Green's function, given by
\begin{equation}
G^r(\omega)=\sum_{n=0}^{\infty} e^{-\gamma^2} \frac{\gamma^{2n}}{n!}
\frac{1}{\omega - \tilde\varepsilon_c + n\omega_p +{\rm i}\eta } ,
\label{eq:exactGr}
\end{equation}
with $\gamma^2=({\gamma_0}/{\omega_p})^2={D}/{\omega_p}$ and the renormalized core level
$\tilde\varepsilon_c = \varepsilon_c + D = \varepsilon_c + \gamma^2 \omega_p$.
The finite temperatures solution is obtained from the prescription given in
Ref.~[\onlinecite{Ness:2006}].

\subsection{Feynman diagrams for the self-energy}
\label{sec:diag}

\begin{figure}
\centering
  (a) \includegraphics[width=3cm]{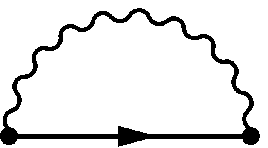}
  (b) \includegraphics[width=3cm]{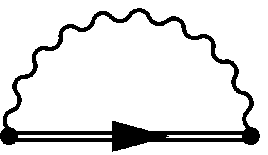}

  (c) \includegraphics[width=32mm]{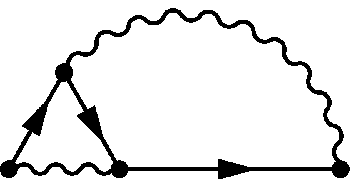}
  (d) \includegraphics[width=32mm]{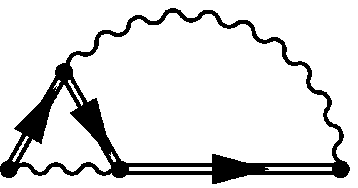}
  \caption{Different levels of approximation for the one-particle
    self-energy $\Sigma=\Sigma^{(1)}+\Sigma^{(2)}$ within the plasmon model.  First order
    diagrams: (a) $\Sigma^{(1)}=G_0 W_p$ with $G_0$ being the bare
    core-electron Green's function; (b) $\Sigma^{(1)}=G W_p$ for
    self-consistent calculations.  Second order diagrams with vertex
    corrections (c) $\Sigma^{(2)}=G \Gamma_{(1)}^{GW} W_p$ for non
    self-consistent calculations; (d) $\Sigma^{(2)}=G \Gamma_{(1)}^{\rm SC}
    W_p$ for the full self-consistent calculations (see Appendix
    \ref{subsec:L1}).}
\label{fig:diagrams}
\end{figure}

The Hamiltonian for the S-model given by Eq.~(\ref{eq:Smodel}) is
effectively a single electron coupled to a single-boson-mode model
similar to the model we studied for an electron-phonon coupled system
in Refs.[\onlinecite{Dash:2010,Dash:2011}]. We can then use the NEGF code we have
developed to study the electronic properties of the S-model for
different levels of approximation for the corresponding
self-energies. In the Feynman diagram language, these are given in
Figure \ref{fig:diagrams} and correspond to (a) non-self-consistent
calculations for the self-energy $\Sigma=G_0 W_p$, where $G_0$ is the
core-electron bare Green's function and $W_p$ is the plasmon
propagator given in Eq.~(\ref{eq:boson_propagator}); (b)
self-consistent calculations for the core electron Green's function
$\Sigma=G W_p$; and to vertex corrections taken at the $\Gamma_{(1)}$ level 
of approximation for (e) non-self-consistent calculations 
$\Sigma=G \Gamma_{(1)}^{GW} W_p$ with $G$ and $\Gamma_{(1)}^{GW}$
taken at the $GW_p$ level 
and (f) fully self-consistent $\Sigma=G \Gamma_{(1)}^{\rm SC} W_p$
calculations.

Our NEGF code, presented in Ref.~[\onlinecite{Dash:2010}] is versatile. It 
was originally developed to
deal with an electron-phonon coupled system in contact with two
electron reservoirs each at their own equilibrium. But the code can deal with
any model Hamiltonian of electron-boson coupled systems.
In the following we
use this code and we consider the whole system at equilibrium, 
and at zero or finite temperature.
As explained above, the exact solution of the S-model exists only for
the equilibrium condition.
 
Additionally we use an extremely small coupling
constant to the reservoirs in order to introduce a finite but
very small broadening in the spectral features of the S-model
Hamiltonian Eq.~(\ref{eq:Smodel}) in a simple way ($\eta$ has a tiny
but finite numerical value). 
The details for the calculations of
the different NEGF, at equilibrium and out of equilibrium, are given
in Ref.[\onlinecite{Dash:2010}].

{ In Ref.~[\onlinecite{Dash:2010}], we discussed the first and
  second-order
  diagrams for the electron-phonon interaction---topologically speaking,
  this will look similar to the $GW$-like self-energy diagrams we consider
  here (Fig.~\ref{fig:diagrams}), however there the boson line is the phonon 
 propagator and not the screened Coulomb interaction $W$ with which we
 are concerned here.
Furthermore the parameters of the core electron-plasmon coupled system are given here
by a single physical quantity: the electron density (see Table \ref{table:param}). }

\subsection{Results}
\label{sec:results}

Within our model, all the characteristics of the plasmon are determined
by a single parameter: the electron density, or equivalently by the
Wigner-Seitz radius $r_S$. There is then only one other parameter left: the
energy level $\varepsilon_c$ of the core electron, which we take as being
located one atomic unit of energy below the Fermi level $\varepsilon_F$ of the
different systems we consider.

\begin{table}
\begin{tabular}{|c||c|c|c|c|}
\hline
$r_S$ 		& 5.0 		& 4.0 		& 3.0 		& 2.0  		\\ \hline
$n$		& 0.00191	& 0.00373	& 0.00884	& 0.02984 	\\
$\varepsilon_F$	& 0.0737	& 0.1151	& 0.2046	& 0.4604   	\\
$\omega_p$	& 0.1549	& 0.2165	& 0.3333	& 0.6124   	\\
$\omega^0_p$	& 2.103		& 1.881		& 1.629		& 1.330   	\\ \hline
$D$		& 0.34046	& 0.31186	& 0.27789	& 0.23523   	\\
$\gamma_0$		& 0.22966	& 0.25985	& 0.30435	& 0.37953   	\\
$\gamma_0/\omega_p$	& 1.48		& 1.20		& 0.91		& 0.62   	\\ \hline
\end{tabular}
\caption{Values (in atomic units) of the different relevant parameters,
electron density $n$, Fermi energy $\varepsilon_F$, plasmon energy $\omega_p$,
electron-plasmon coupling constant $\gamma_0$ and relaxation energy $D$ for
different values of $r_S$.}
\label{table:param}
\end{table}

Table \ref{table:param} contains the values of the different relevant
parameters for four different values of $r_S$.  The high-density limit
($r_S=2$) corresponds to a medium electron-plasmon coupling, while the
low-density limit ($r_S=5$) corresponds to a very strong
electron-plasmon coupling.

Below, and in Figs.~\ref{fig:Aw_rs2}--\ref{fig:Aw_rs4_Tdep_lesspts},
we show results for the spectral function $A(\omega)= -{\rm
  i}(G^r(\omega)-G^a(\omega))/2$ calculated at equilibrium for two
values of $r_S$ (medium coupling $r_S=2$, strong coupling $r_S=4$) at
zero temperature and at a finite temperature.  We compare the exact
results Eq.~\eqref{eq:exactGr} for the spectral function with
the results obtained from the diagrammatic expansion of the
self-energy and the vertex function shown in Fig.~\ref{fig:diagrams}.

\subsubsection{Exact results}
\label{sec:Exact_results}

\begin{figure}
  \includegraphics[width=\columnwidth]{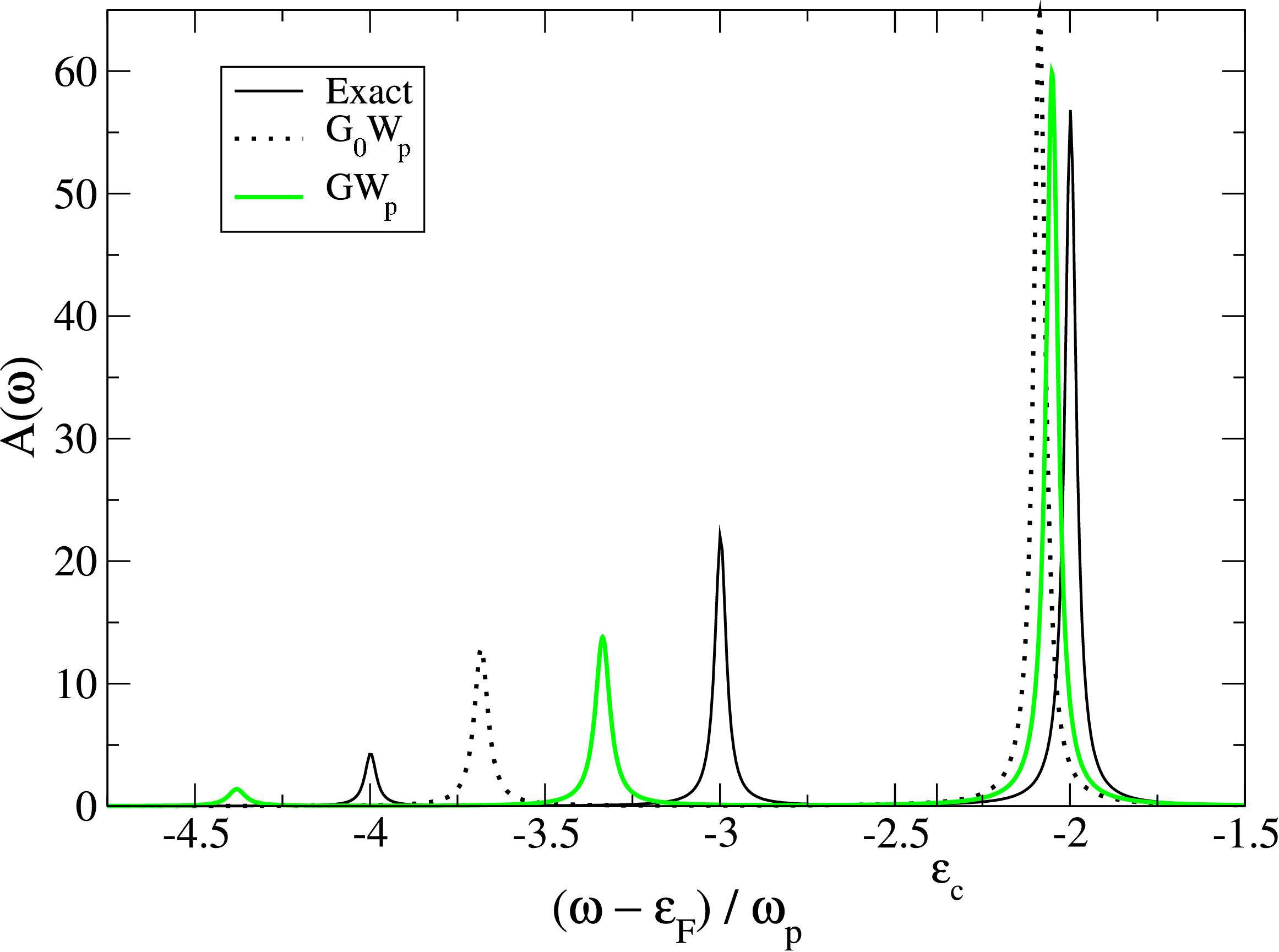}

  \vspace{11mm}

  \includegraphics[width=\columnwidth]{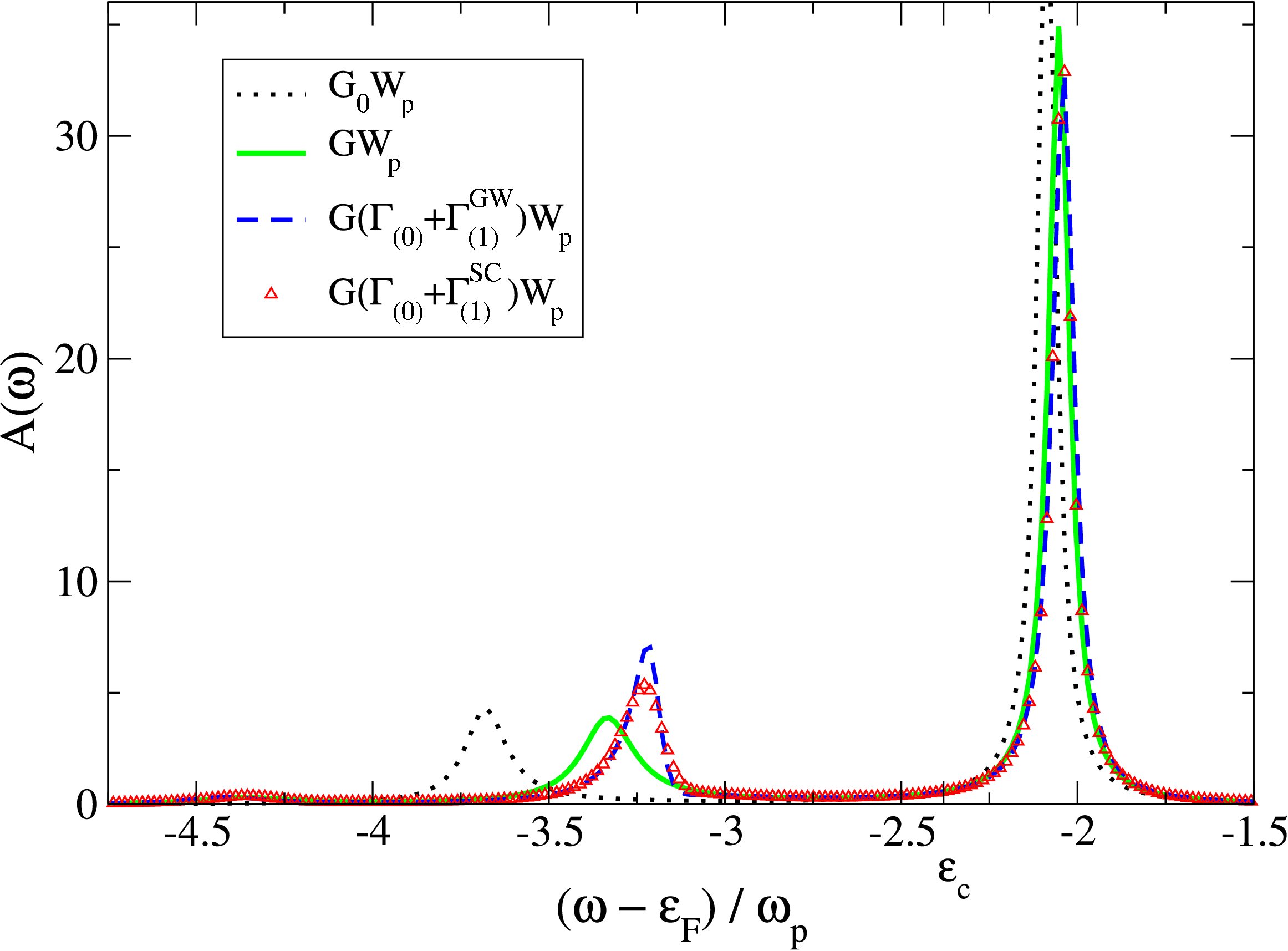}
  \caption{(Color online) Zero-temperature equilibrium spectral
    functions $A(\omega)$ for the for the high-density limit
    with $r_S=2$, corresponding to medium core electron-plasmon
    coupling $\gamma_0/\omega_p=0.62$.  Top panel: Exact results and
    $GW$ calculations with and without self-consistency $\Sigma=G W_p$, $G_0 W_p$.  
    Bottom panel: Results for different levels of
    approximation for the self-energy $\Sigma=G_0 W_p$, $G W_p$, 
      $G (\Gamma_{(0)}+\Gamma_{(1)}^{GW}) W_p$ and $G (\Gamma_{(0)}+\Gamma_{(1)}^{\rm
      SC}) W_p$ (see Fig.~\ref{fig:diagrams}) with fewer
    grid points ($N_\omega=1579$), giving an extra broadening.}
  \label{fig:Aw_rs2}
\end{figure}

\begin{figure}
  \includegraphics[width=\columnwidth]{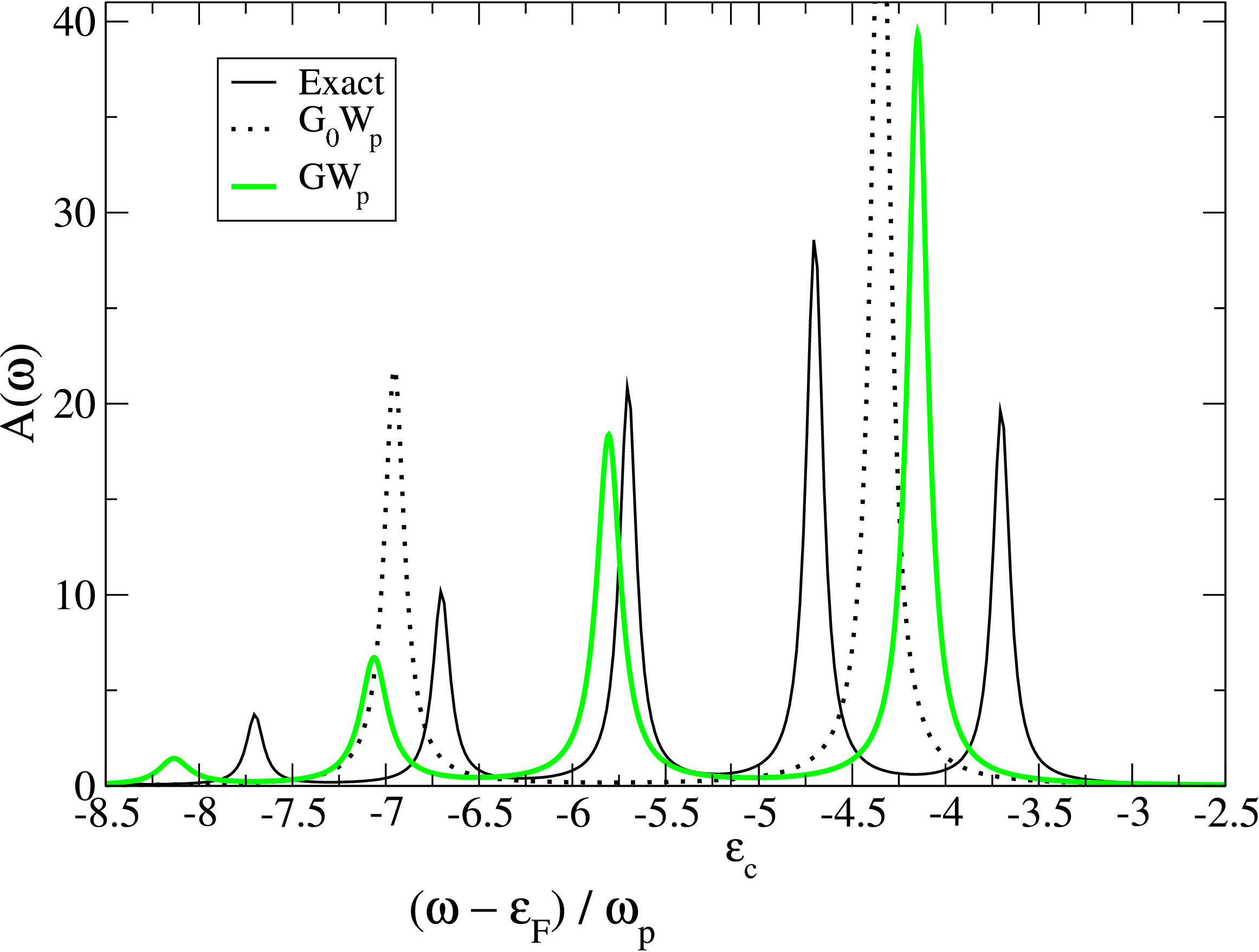}
  
  \vspace{11mm}

  \includegraphics[width=\columnwidth]{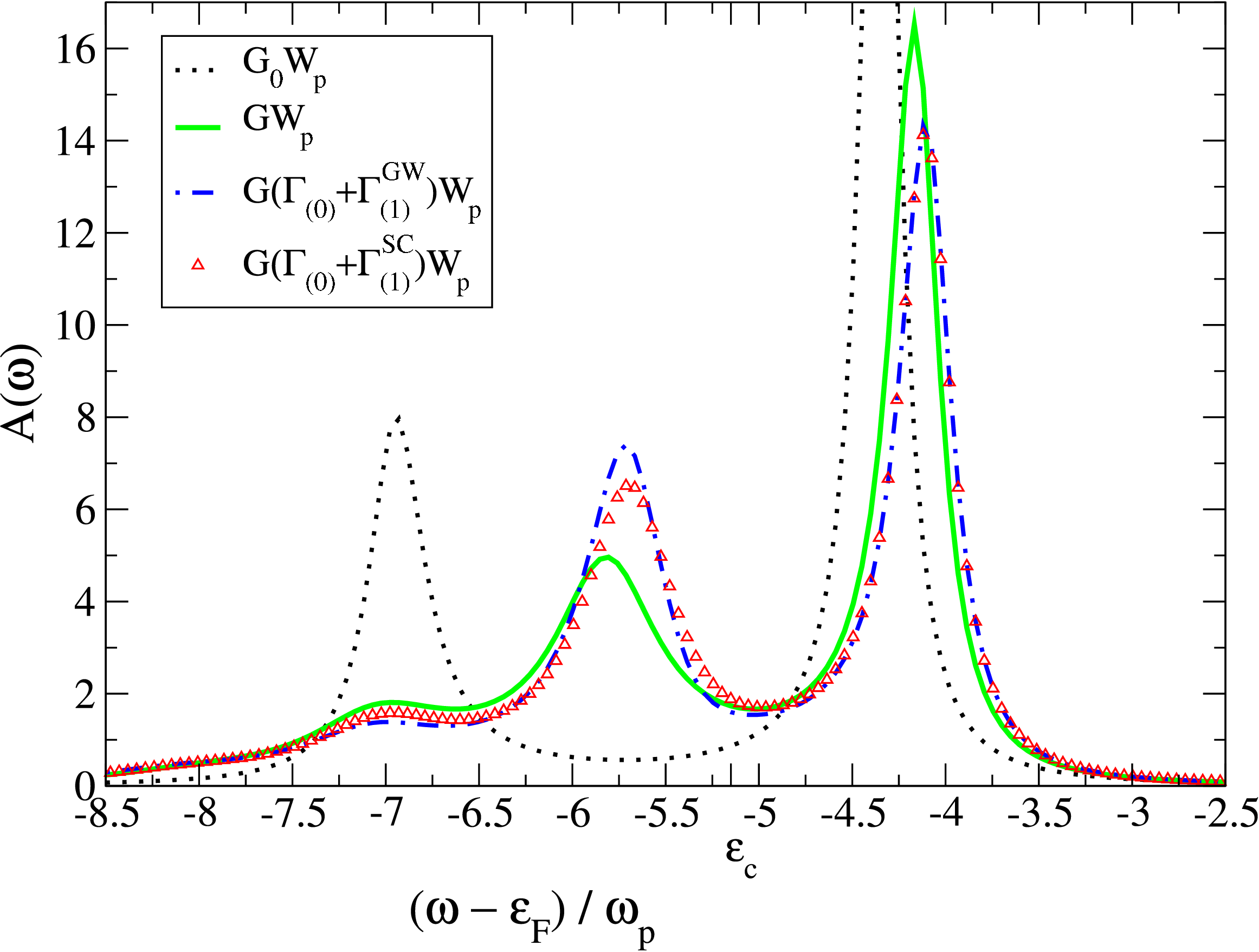}
  \caption{(Color online) Zero-temperature equilibrium spectral
    functions $A(\omega)$ for the low-density limit with
    $r_S=4$, corresponding to very strong core electron-plasmon
    coupling $\gamma_0/\omega_p=1.20$. Top panel: Exact results and
    $GW$ calculations for the different self-energies 
	$\Sigma=G_0 W_p$, $G W_p$.  Bottom panel: Results for
    different levels of approximation for the self-energy $\Sigma=G_0
    W_p$, $G W_p$, $G (\Gamma_{(0)}+\Gamma_{(1)}^{GW}) W_p$ and $G (\Gamma_{(0)}+\Gamma_{(1)}^{\rm
      SC}) W_p$ (see Fig.~\ref{fig:diagrams}) with fewer
    grid points ($N_\omega=1579$), giving an extra broadening,
    in comparison to the top panel.}
  \label{fig:Aw_rs4}
\end{figure}

The exact spectral function, calculated from the expression for the
Green's function given in Eq.~(\ref{eq:exactGr}), is shown as a solid
black line in Figs.~\ref{fig:Aw_rs2} and \ref{fig:Aw_rs4}.  A
broadening equal to the broadening of our NEGF calculations has been applied.
Fig.~\ref{fig:Aw_rs2} shows the zero-temperature results for the
high-density electron gas ($r_S=2$).  The exact result provided by
Eq.~(\ref{eq:exactGr}) (solid black line) gives a spectral function
with a peak localized at the renormalized core level
$\tilde\varepsilon_c=\varepsilon_c+D$, and plasmon side-band peaks at
$\tilde\varepsilon_c-n\omega_p$ ($n \ge 1$) corresponding to plasmon
emission.  The peaks are hence separated by the plasmon energy
$\omega_p$.  In terms of amplitude, the main peak is that at
$\tilde\varepsilon_c$ in the limit of weak to medium/strong
electron-plasmon coupling, i.e.\ where $\gamma_0/\omega_p \le 1$, and
so for which $\gamma_0/\omega_p \le 1$.

Fig.~\ref{fig:Aw_rs4} shows the zero-temperature results for stronger
coupling, $r_S= 4$ and $\gamma_0/\omega_p > 1$.  Now the renormalized
core level $\tilde\varepsilon_C$ has shifted to $\omega/\omega_p\sim 1.5$, while the
spectral weight is shifted towards lower energies and the main peak is
now the plasmon side-band peak at around $\omega/\omega_p$
\cite{Mahan:1990}.

\subsubsection{Diagrammatic expansion results}
\label{sec:calculated_results}

The main differences between the exact result and the diagrammatic
expansions of the self-energies and of the vertex functions (as
represented in Fig.~\ref{fig:diagrams}) are as follows:

First, let us discuss the results for the spectral functions in
the high-density limit ($r_S=2$) for which the electron-plasmon
coupling is medium $\gamma_0/\omega_p=0.62$.

The non-self-consistent $GW$ calculations (i.e.\ $\Sigma=G_0 W_p$,
Fig.~\ref{fig:diagrams}(a), dotted black lines in
Fig.~\ref{fig:Aw_rs2}) generate only two
peaks, the renormalized core level with one plasmon side-band peak, as
expected. However the positions of those two peaks are incorrect.

The self-consistent $GW$ calculations (i.e. $\Sigma=G W_p$,
Fig.~\ref{fig:diagrams}(b), solid green lines in
Figs.~\ref{fig:Aw_rs2} and \ref{fig:Aw_rs4}) generate the correct
series of plasmon side-band peaks. 
However the corresponding relaxation energy $D$ is
too small and the energy position of the first plasmon side-band peak
is too low.  It should be
noticed however that the energy separation between the plasmon
side-band peaks is correctly reproduced, i.e. equal to $\omega_p$.

For the low-density limit ($r_S=4$) for which the electron-plasmon
coupling is very strong $\gamma_0/\omega_p=1.20$, the $GW$
calculations poorly describe the exact spectral density. The
self-consistent $GW$ calculations generate the correct series of peaks
but with a completely wrong weight distribution.  This is unsurprising
since the $GW$ approach corresponds to a partial resummation of the
diagrams, and does not include all other relevant diagrams necessary to
deal with the very strong regime.

The lowest-order vertex corrections to the self-energy
(Figs.~\ref{fig:diagrams}(d) and \ref{fig:diagrams}(e), blue dashed
lines and red triangles in Figs.~\ref{fig:Aw_rs2} and
\ref{fig:Aw_rs4}) introduce modifications of the peak positions. They
generate a slightly better relaxation energy $D$ and a shift of the
side-band peaks towards the renormalized electron core level
(Figs.~\ref{fig:Aw_rs2} and \ref{fig:Aw_rs4}, bottom panels). Vertex
corrections globally improve the spectral information towards 
better overall agreement with the exact results.  However, the lowest-order
vertex correction expansion $\Gamma_{(0)}+\Gamma_{(1)}$ (see Appendix \ref{app:lowestorder_exp})
is still not
sufficiently good to qualitatively reproduce the exact spectral
functions in the limit of very strong electron-plasmon coupling.

The fully self-consistent calculations with $G \Gamma_{(1)}^{\rm SC} W_p$ seem
to only marginally affect the lineshape of the plasmon side-band peaks in 
comparison to their non  self-consistent counterpart.

Note that a fine analysis of the comparison between the exact results
and the diagrammatic perturbation results with vertex correction is
difficult to perform in
Figs.~\ref{fig:Aw_rs2} and \ref{fig:Aw_rs4}, as the calculations were
done for different numbers of $\omega$-grid points $N_\omega$.  
It was necessary
to perform the calculations in that way because the vertex corrections
scale as $N_\omega^3$ as shown in Ref.~[\onlinecite{Dash:2010}].
Therefore we have performed the corresponding calculations with a
lower number of points $N_\omega=1579$ for the bottom panels of
Figs.~\ref{fig:Aw_rs2} and \ref{fig:Aw_rs4}, instead of
$N_\omega=16385$ points for the top panels, in order to have tractable
computational costs.  Our NEGF code works with a finite broadening
related to the number of grid points to deal with sharply peaked
and/or discontinuous functions, hence the different lineshape in the
spectral functions in the top and bottom panels of
Figs.~\ref{fig:Aw_rs2} and \ref{fig:Aw_rs4} respectively.  This
numerical extra broadening affects only the width of the peaks and the
global amplitude of the spectral functions, though all spectral functions
are always normalized. There is no major problem with
the spectral information contained in $A(\omega)$.  We have discussed in detail the
effects of this extra broadening in Ref.~[\onlinecite{Dash:2010}].

In addition, we want to add that our results confirm those obtained in
earlier studies, see for example
Refs.~[\onlinecite{Langreth:1970,Minnhagen:1975,Verdozzi:1995,Shirley:1996}].
However our self-consistent scheme for calculating the
second-order diagrams by starting with the $GW$-like Green's function
allows us to avoid the problem of negative spectral densities (at
least within the range of parameters we have explored) that were
obtained in Refs.~[\onlinecite{Minnhagen:1974,Minnhagen:1975,Verdozzi:1995}].

\subsubsection{Finite temperatures}
\label{sec:Finite_temperatures}

\begin{figure}
  \includegraphics[width=\columnwidth]{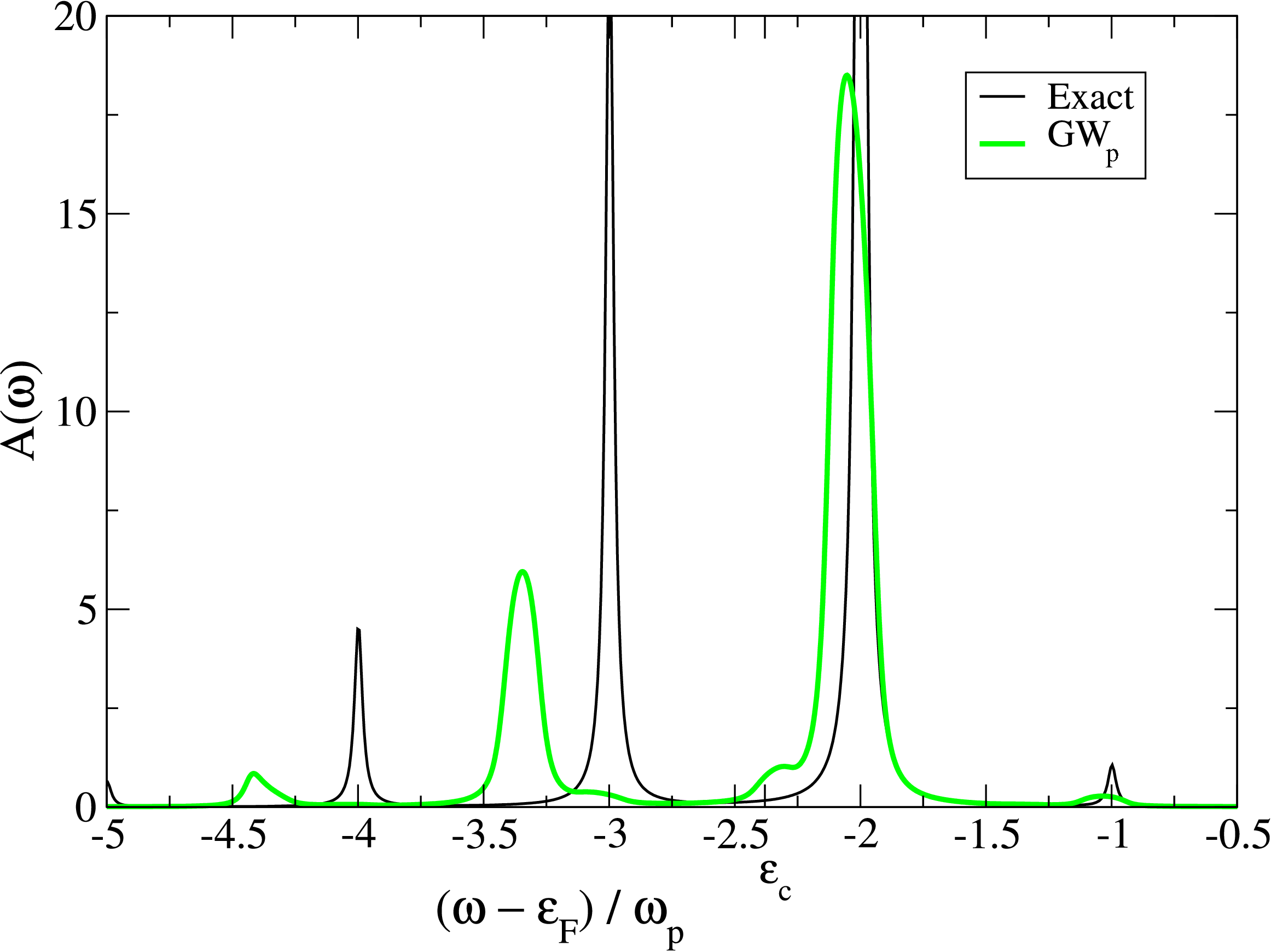}

  \vspace{11mm}

  \includegraphics[width=\columnwidth]{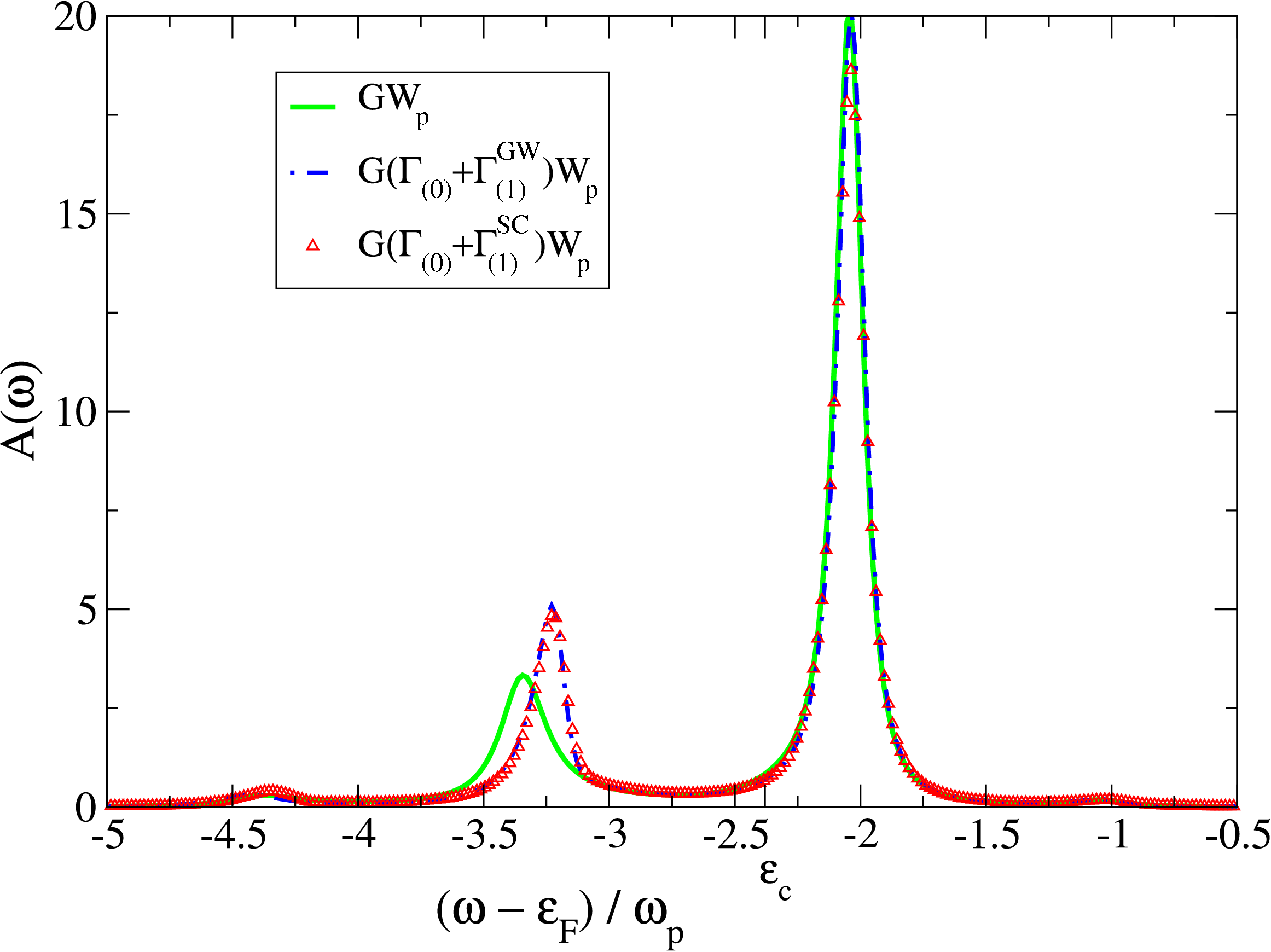}
  \caption{(Color online) Finite-temperature equilibrium spectral
    functions $A(\omega)$ for the high-density electron gas with
    $r_S=2$ and a finite temperature $kT=0.2$ corresponding to
    $\omega_p/kT=3.062$.  Top panel: Exact results and calculations
    for different self-energies $\Sigma=G W_p$ and $G W^{(2){\rm
        SC}}_p$.  Bottom panel: Results for different self-energies
    $\Sigma=G W_p$, $G (\Gamma_{(0)}+\Gamma_{(1)}^{GW}) W_p$ and $G (\Gamma_{(0)}+\Gamma_{(1)}^{\rm
      SC}) W_p$ (see Fig.~\ref{fig:diagrams}) with fewer
    grid points ($N_\omega=1579$), giving an extra broadening.}
  \label{fig:Aw_rs2_Tdep}
\end{figure}

\begin{figure}
  \includegraphics[width=\columnwidth]{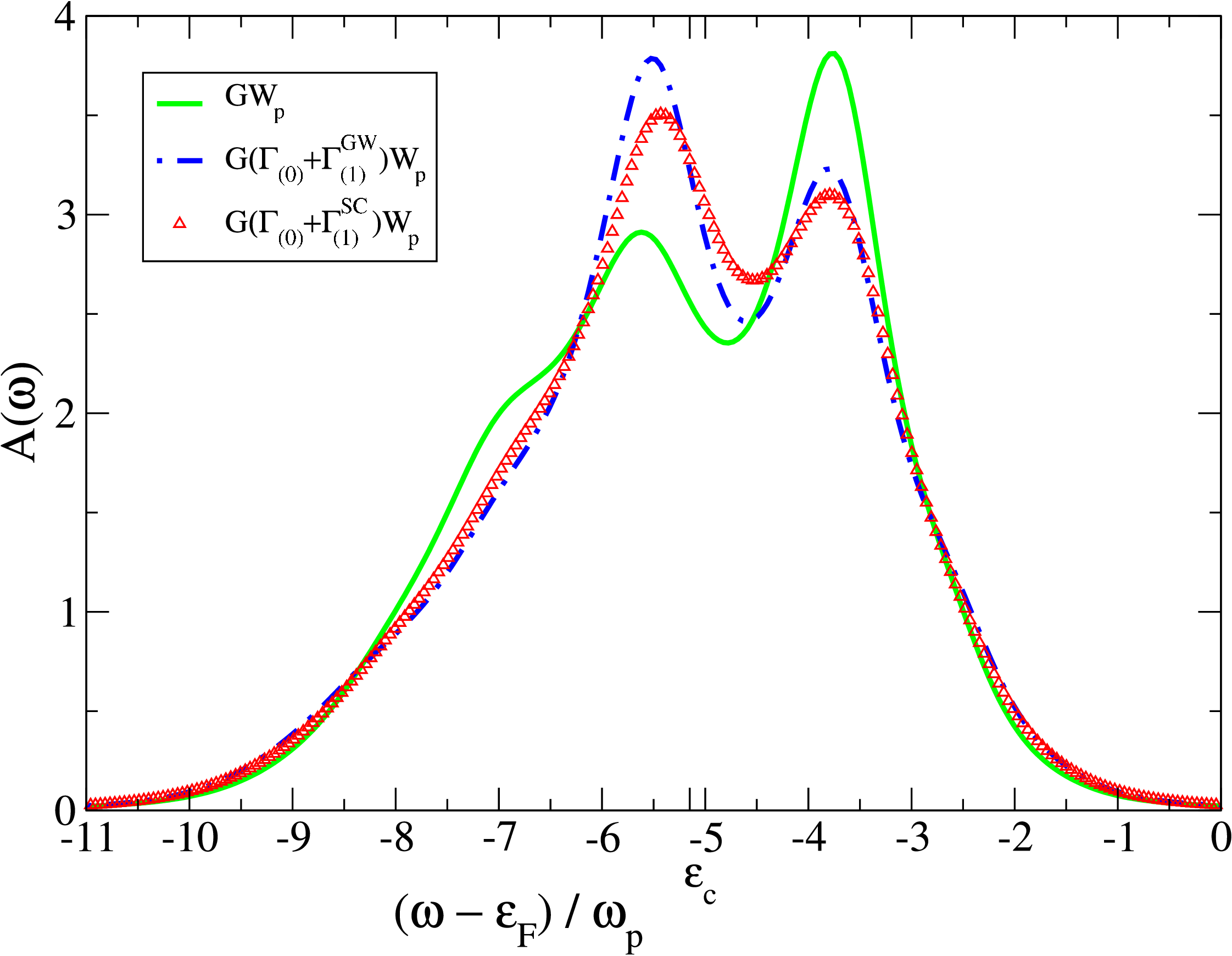}
  \caption{(Color online) Finite-temperature equilibrium spectral
    functions $A(\omega)$ for the low-density electron gas with
    $r_S=4$ and a finite temperature $kT=0.2$ (corresponding to
    $\omega_p/kT=1.083$) and with fewer grid points
    $N_\omega=1579$.  Calculations for different self-energies
    $\Sigma=G W_p$, $G W^{(2){\rm SC}}_p$, $G (\Gamma_{(0)}+\Gamma_{(1)}^{GW}) W_p$
    and $G (\Gamma_{(0)}+\Gamma_{(1)}^{\rm SC}) W_p$ (see Fig.~\ref{fig:diagrams})
    are shown.}
  \label{fig:Aw_rs4_Tdep_lesspts}
\end{figure}

For finite temperatures, the exact result provided by
Eq.~(\ref{eq:exactGr}) can be generalized from a thermodynamical
average over the boson statistics within a canonical
ensemble\cite{Mahan:1990,Ness:2006}.  In addition to the peaks at
$\tilde\varepsilon_c-n\omega_p$ ($n \ge 0$), one also sees spectral
information at $\tilde\varepsilon_c+n\omega_p$ ($n \ge 1$) which
corresponds to absorption of the thermally populated plasmons, as
shown in Figure \ref{fig:Aw_rs2_Tdep}.

The results for the spectral functions obtained from the diagrammatic
expansion of the self-energy and of the vertex functions as shown in
Fig.~\ref{fig:diagrams} are shown in Figs.~\ref{fig:Aw_rs2_Tdep} and
\ref{fig:Aw_rs4_Tdep_lesspts}.  Qualitatively we obtain similar
effects of the second-order diagrams on the spectral functions as in
the case of zero temperature.  Note that however, for finite
temperatures, the dependence of the lineshape upon the extra
broadening related to the number of $\omega$-grid points is much less
important, since the thermal broadening is dominating.  
In Fig.~\ref{fig:Aw_rs2_Tdep} we see that, as for the zero-temperature
case, the self-consistent $GW_p$ calculations generate the correct series of 
peaks with the plasmon emission sideband peaks again
appearing at too low energies. However the new plasmon absorption peak just
above the main peak is almost at the correct energy position.

{ We do not yet have an accurate explanation for the tiny
  shoulder-like feature around the Fermi level in the top panel of
  Fig.~\ref{fig:Aw_rs2_Tdep}.  However, this feature is related to
  plasmon absorption processes since at the chosen temperature the
  plasmon mode can be thermally populated.  Nonetheless, it is clear that
  the feature disappears when performing the calculations with an extra
  broadening (i.e.\ introducing an effective finite lifetime for the
  plasmon mode).}

When we consider the strong coupling case, shown in
Fig.~\ref{fig:Aw_rs4_Tdep_lesspts}, we find that for all levels of
approximation the lineshape is strongly broadened, washing out most of
the features.  

We can conclude that, within the limit of the S-model and
for both the zero-temperature and
finite-temperature cases, the various $GW$ approximations are much
more accurate for the high-density regime.  For the low-density
electron gas both the $GW$ peak positions and lineshapes are poor in
comparison to the exact results, although the separation between the
plasmon sideband peaks is correctly reproduced.

\subsection{Spectral function of pure jellium and vertex corrections}
\label{martin-jellium}

\begin{figure}
  \includegraphics[width=0.92\columnwidth]{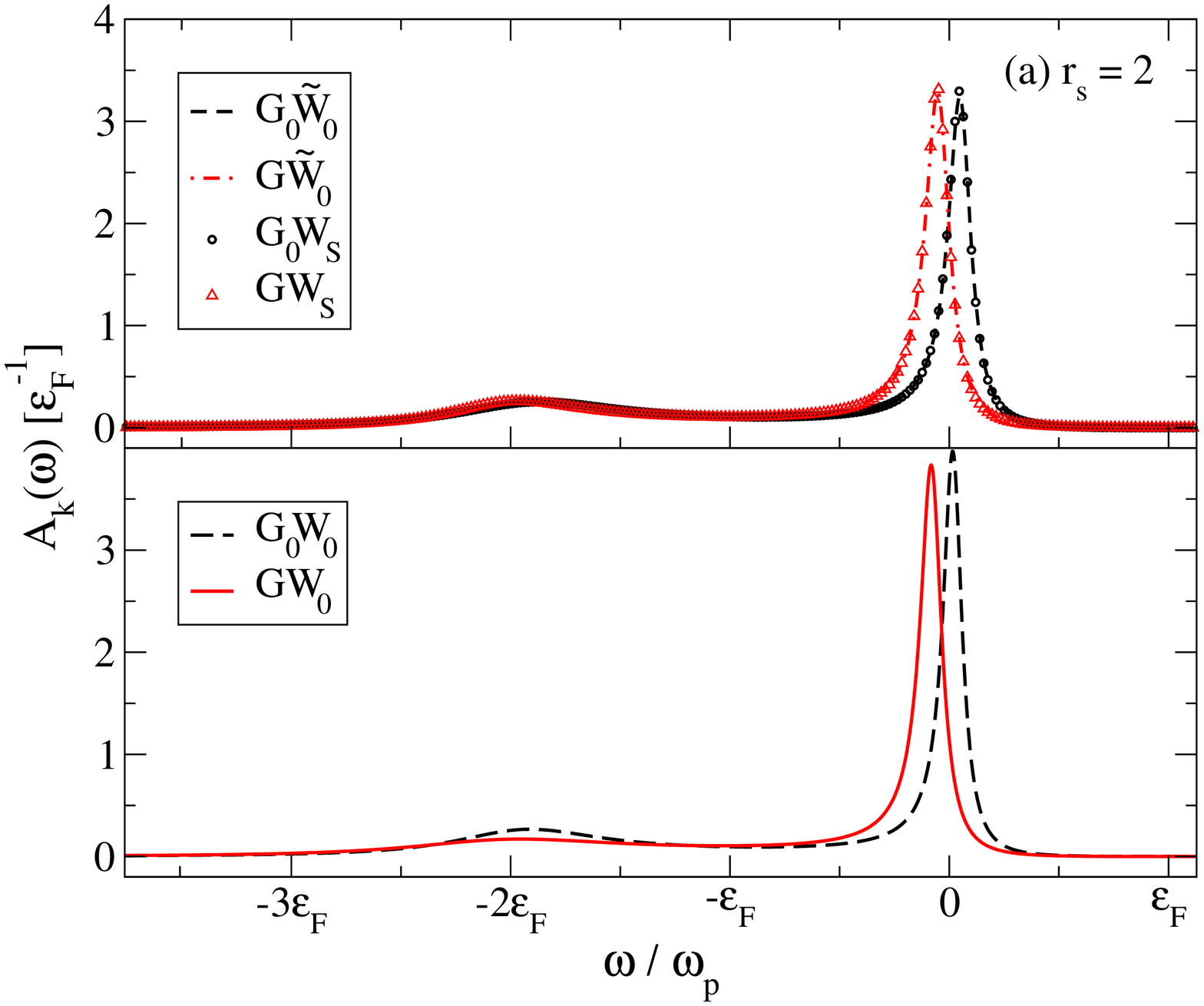}

  \vspace{2mm}

  \includegraphics[width=0.92\columnwidth]{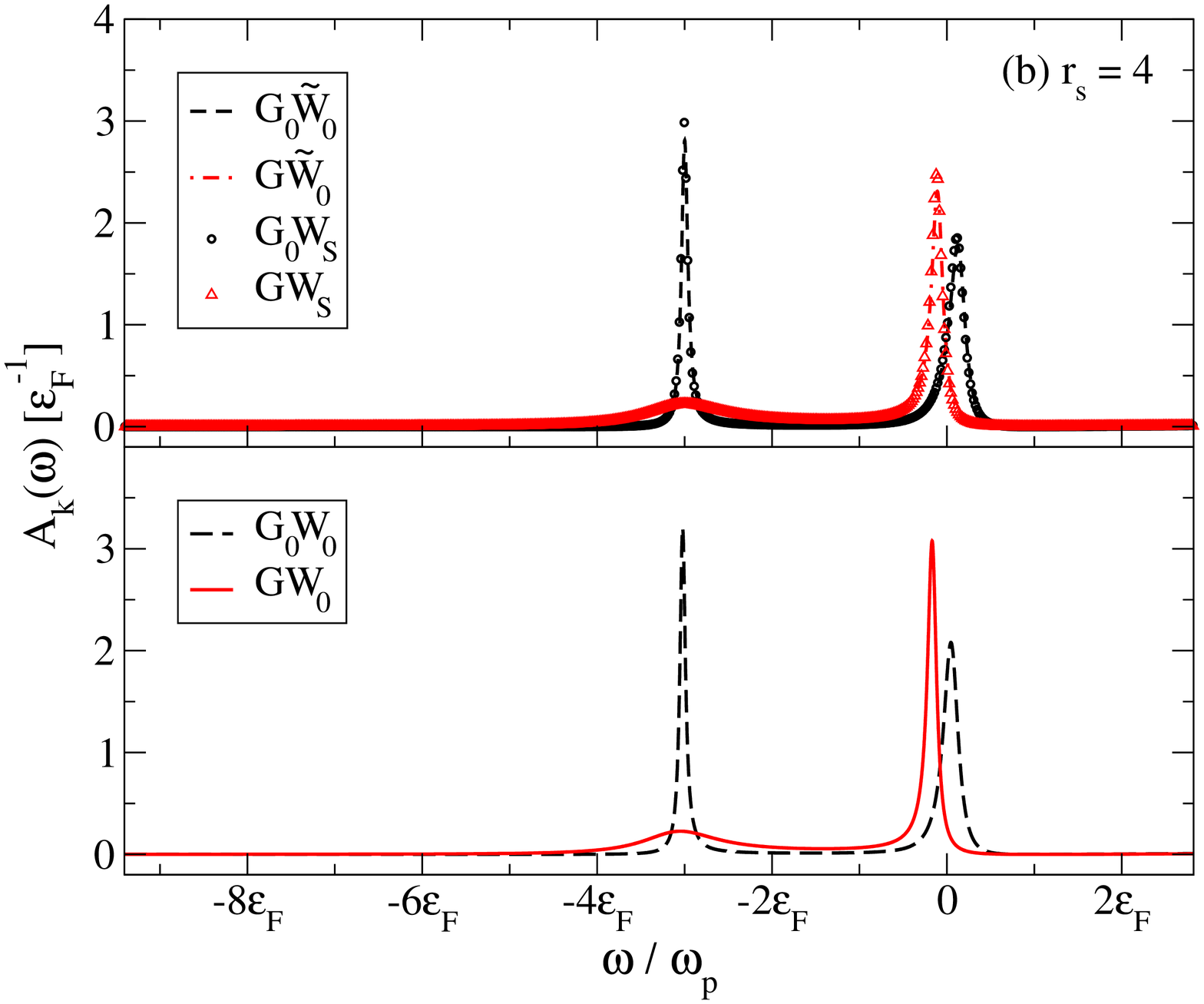}
  \caption{(Color online) Spectral function $A_k(\omega)$ at $k=0$ and
    for (a)
    $r_\mathrm{S}=2$ and (b) $r_\mathrm{S}=4$ for the pure
    jellium model. Each set of curves shows a one-shot calculation
    for $G_0$ 
    (black data) and self-consistent iterations for $G$ (red data).  
    The bottom panel of plots (a) and (b) show standard $G_0W_0$ and $GW_0$, 
    the top
    panels show $G_0\tilde{W}_0$ and $G\tilde{W}_0$ with a local
    vertex correction in $\tilde{W}_0$ as described in
    Ref.~[\onlinecite{DelSole:1994}]. $G_0W_{\mathrm{S}}$ and
    $GW_\mathrm{S}$ refer to a momentum-dependent vertex correction
    in $W_{\mathrm{S}}$
    defined in Ref.~[\onlinecite{Shirley:1996}] to approximate the
    exact $W$ of jellium.  
    The results including the different vertex
    corrections are indicated with a line (for $\tilde{W}_0$) and
    a symbol (for $W_{\mathrm{S}}$). All
    chemical potentials are aligned at the Fermi energy $\epsilon_F$
    of the noninteracting gas. The positive (negative) deviation of 
    the main quasiparticle peak from the origin indicates a narrowing
    (broadening) of the occupied bandwidth. None of these approximations 
    provides more than one plasmon satellite in contrast to the
    expected exact result.}
  \label{fig:martin_jellium_plot}
\end{figure}

In this section, we compare different approximations for the vertex 
corrections for another model system: the pure jellium model 
(without a distinct core level). The spectral functions in 
this system are evaluated in the zero-temperature limit within conventional
Green's functions calculations \cite{Stankovski:unpub}.

It is expected from the original work of Hedin \emph{et
  al.}~\cite{Hedin:1969} and also of Shirley\cite{Shirley:1996} that the
exact spectral function of pure jellium should show several plasmon
resonances below the main quasiparticle peak.  However, we do not
observe any such peaks (see Fig.~\ref{fig:martin_jellium_plot})
when iterating the Green's function to self-consistency within the $GW$
approximation, nor when we use model vertex corrections \cite{DelSole:1994,Shirley:1996}. 
These vertex corrections were however
supposed to provide an exact description of screened Coulomb
interaction $W$ for the jellium model.

Any self-consistent iteration has the effect of broadening the
occupied bandwidth (a feature which is known to be unphysical) as
evidenced by the shift in the main quasiparticle peak at the bottom of
the band seen in Fig.~\ref{fig:martin_jellium_plot}. The model vertex
corrections tested do not remedy this behavior, nor do they lead to
any multi-plasmon resonances.  We consider two different models for the
vertex corrections: Firstly, a strictly local vertex correction applied in
the screening, annotated $\tilde{W}_0$ and modelled directly by the LDA
exchange-correlation kernel as described by Del Sole \emph{et
  al.}~\cite{DelSole:1994}.  Secondly, the other vertex correction
incorporates a momentum-dependent local field factor modelled on exact
quantum Monte Carlo results for jellium, as described by
Shirley~\cite{Shirley:1996} (annotated $W_{\mathrm{S}}$).

In general, the difference between the two different types (static
vs. $q$-dependent) of vertex corrections implemented is practically
negligible in the spectral functions. This shows that the screened
interaction can be very insensitive to the exact type of vertex
correction used, in contrast to the self-energy. With a
self-consistent calculation, we also observe the broadening of
spectral peaks previously noted in
Refs.~[\onlinecite{vonBarth1996,Holm1998}].

This also indicates that the explicit evaluation of the second-order
diagrammatic vertex correction, ${\Gamma}_{(1)}$, is imperative in
order to capture the higher-order plasma sattelites in a metallic
system, and in corresponding models with a coupling to a core
state as shown in the previous section. 
This finding is fully consistent with the previous work of
Shirley \cite{Shirley:1996} where the vertex function ${\Gamma}_{(1)}$
was approximately evaluated within the zero-temperature formalism.

\section{Conclusions}
\label{ccl}

We have formally expressed the Hedin's $GW$ equations on the
Keldysh time-loop contour. This implies that within our
formalism one can now deal with full non-equilibrium conditions for
fully interacting electron systems. The equilibrium properties of the
system are obtainable from our formalism as a special case of the more
general non-equilibrium conditions.

We have considered in particular the lowest-order expansions of the
electron self-energy $\Sigma$ and of the vertex function $\Gamma$, and
compare our results with previous work.  We have then used our
formalism to study a simple model of an electron core level coupled to
a plasmon mode for which exact results for the spectral function are
available (i.e.\ the S-model).  We have compared our lowest-order
expansions of the electron self-energy and of the vertex function with
the exact results, considering the second-order diagrams in terms of
the plasmon propagator $W_p$.

We have shown that self-consistent $GW$-based approximations (with or without vertex
corrections) provide a good approximation to the exact results in the
limit of weak to medium electron-plasmon coupling (i.e.\ high
electron-density limit) both at zero and finite temperatures.
Non self-consistent $G_0W_p$ calculations do not reproduce the complete
series of plasmon sattelites.
However the $GW$ based approximations
perform quite poorly in the strong-coupling limit (i.e.\ low
electron-density limit).  
Vertex corrections generally re-adjust the peak positions (the 
relaxation energy responsible for the renormalization of the core level 
as well as the plasmon side-band peaks) towards the correct
result.

Furthermore we have also analyzed the spectral functions obtained
from conventional equilibrium $GW$ calculations for the pure jellium model
and using different approximation for the vertex corrections in $W$. 
The corresponding results confirm that the explicit second order diagrams for
the vertex corrections are needed to obtain the full series of plasmon
side-band resonances.

In appendix \ref{app:timeorderedGFs}, we have also addressed an important 
issue about the Dyson-like equation for the time-ordered Green's function in the energy
represention. We have shown that there is a difference
between Dyson equation for the Green's function obtained at zero-temperature 
and at finite temperature, as already pointed out in Ref.~[\onlinecite{Hedin:1969}]. 
We have shown that at finite
temperature there are extra terms in the Dyson equation of the
time-ordered Green's function. These terms are obtained rigorously from the Keldysh time-loop 
formalism we derived at equilibrium, while they were introduced {\it ad hoc} 
by Hedin and Lundqvist \cite{Hedin:1969} to recover an exact result.

Finally, we have studied in this paper models of interacting electron systems,
but we believe that our theoretical approach is well-suited for applications towards
more realistic physical systems, such as the one-dimensional plasmon modes 
recently observed in an atomic-scale metal wire deposited on a surface \cite{Nagao:2006}.

\begin{acknowledgments}
We gratefully acknowledge Pablo Garc\'ia Gonz\'alez for useful
discussions, comments,
and the use of a version of his jellium code.
This work was funded in part by the European Community's Seventh
Framework Programme (FP7/2007-2013) under grant agreement no 211956
(ETSF e-I3 grant).
\end{acknowledgments}

\appendix

\section{Relationship between the different Green's functions and self-energies}
\label{app:GFSEdefinitions}

The relations between the different components of the Green's
functions and self-energies on the Keldysh time-loop contour are given
as usual, with $X^{\eta_1 \eta_2}(12)\equiv G^{\eta_1 \eta_2}(12)$ or
$\Sigma^{\eta_1 \eta_2}(12)$.
\begin{equation}
\begin{split}
X^r = X^{++}-X^{+-} &= X^{-+}-X^{--} \\
X^a = X^{++}-X^{-+} &= X^{+-}-X^{--} \\
X^{++}+X^{--} &= X^{+-}+X^{-+}     \\ 
X^{-+}-X^{+-} &= X^r-X^a
\end{split}
\label{eq:app_gendef}
\end{equation}

The usual lesser and greater projections are defined respectively as
$X^< \equiv X^{+-}$ and $X^> \equiv X^{-+}$, and the usual
time-ordered (anti-time-ordered) as $X^t=X^{++}$ ($X^{\tilde
  t}=X^{--}$).

\section{Rules for analytical continuation}
\label{app:analyticalcontinuation}

For the following products $P_{(i)}(\tau,\tau')$ on the time-loop contour $C_K$,
\begin{equation}
\label{eq:app_acontinuation_CK}
\begin{split}
P_{(2)}	& =  \int_{C_K} A B   \\
P_{(3)}	& =  \int_{C_K} A B C  \\
P_{(n)}	& =  \int_{C_K} A_1 A_2 ... A_n,  \\ 		\nonumber
\end{split}
\end{equation}
we have the following rules for the different components  $P_{(i)}^x(t,t')$ on the real-time axis:
$(x=r,a,>,<)$
\begin{equation}
\label{eq:app_acontinuation_t}
\begin{split}
P_{(2)}^\gtrless	& =  \int_t A^r B^\gtrless + A^\gtrless B^a   \\ 
P_{(3)}^<	& =  \int_t A^< B^a C^a + A^r B^< C^a + A^r B^r C^<  \\
P_{(n)}^r	& =  \int_t A_1^r A_2^r ... A_n^r  \ , \hspace*{5mm}
P_{(n)}^a	 =  \int_t A_1^a A_2^a ... A_n^a.  	\nonumber
\end{split}
\end{equation}

\section{Lowest order expansion of the vertex function $\Gamma(12;3)$}
\label{app:lowestorder_exp}

\subsection{The $\Gamma_{(0)}$ level of approximation: no vertex corrections}
\label{subsec:noVtx}

In this section, we derive from our general results the more
conventional $GW$ approach used in previous studies on the ground state
properties of molecules, semi-conductors, or on the linear
response or the full non-equilibrium transport properties of nanoscale
systems driven by an applied external voltage
\cite{Strange:2011,Mera:2010,Rostgaard:2010,Spataru:2004,Spataru:2009,Wang:2008,Thygesen:2007,Darancet:2007}.

With no vertex corrections, $\Gamma(12;3)$ is simply given by
$\Gamma_{(0)}(12;3)=\delta(12)\delta(13)$.  Hence the polarizability
$\tilde{P}(12)$ and the electron self-energy $\Sigma(12)$ are
\begin{equation}
\begin{split}
\tilde{P}(12) &= -{\rm i} G(12)\ G(21), \\
\Sigma(12) &= {\rm i} G(12)\ W(21) .
\end{split}
\end{equation}
The different components of the polarizability are then
\begin{equation}
\tilde{P}^\lessgtr(12)=-{\rm i} G^\lessgtr(12)\ G^\gtrless(21) .
\end{equation}
Using Eqs.~(\ref{eq:app_gendef}), we find that the retarded polarizability
is given by
\begin{equation}
\begin{split}
\tilde{P}^r(12) & = -{\rm i} [G(12)\ G(21)]^r \\
& = -{\rm i} G^r(12)\ G^<(21) -{\rm i} G^<(12)\ G^a(21) ,
\end{split}
\end{equation}
and the electron self-energy by
\begin{equation}
\begin{split}
\Sigma^<(12)
&={\rm i} G^<(12)\ W^>(21), \\
\Sigma^r(12)
& = {\rm i} [G(12)\ W(21)]^r \\
& = {\rm i} G^r(12)\ W^<(21) +{\rm i} G^<(12)\ W^a(21) .
\end{split}
\end{equation}
Using the symmetry relations for $W$ and Eqs.~(\ref{eq:app_gendef}), we can 
easily recast the above equations in the following form 
\begin{equation}
\begin{split}
\Sigma^<(12)
&={\rm i} G^<(12)\ W^<(12) \\
\Sigma^r(12)
&={\rm i} G^r(12)\ W^>(12) +{\rm i} G^<(12)\ W^r(12) .
\end{split}
\end{equation}
These expressions for $\Sigma$ and $\tilde{P}$ are just the equivalent of 
Eqs.~(3-8) in Ref.~[\onlinecite{Thygesen:2007}] and are similar to the 
corresponding expressions in Refs.~[\onlinecite{Stan:2009a,Stan:2009b,Rostgaard:2010,Spataru:2004}].

\subsection{The $\Gamma_{(1)}$ level of approximation}
\label{subsec:L1}

With the series expansion $\Gamma(12;3)=\sum_n \Gamma_{(n)}(12;3)$, in
which the index $n$ represents the number of times the screened
Coulomb interaction $W$ appears explicitly in the series, we take for
$\Gamma_{(1)}(12;3)$
\begin{equation}
\Gamma_{(1)}(12;3)=
\int {\rm d}(4567)\ 
\frac{\delta \Sigma(12)}{\delta G(45)}
G(46)  G(75) \Gamma(67;3) ,
\end{equation}
where$\Gamma(67;3)=\Gamma_{(0)}(67;3)=\delta(67)\delta(63)$ and
$\Sigma={\rm i} G W$.  Hence $\Gamma_{(1)}(12;3)={\rm i} W(21)\ G(13)\
G(32)$.

In the following, we derive the part of the electron self-energy 
and the part of the polarizability arising from $\Gamma_{(1)}$ only.
In principle, the full $\Sigma$ and $\tilde{P}$ should be calculated by
using  $\Gamma=\Gamma_{(0)}+\Gamma_{(1)}$.
We find for the electron self-energy (defined on the contour $C_K$):
\begin{equation}
\begin{split}
\Sigma(12) & = {\rm i}
\int {\rm d}(34)\
G(13)\ \Gamma_{(1)}(32;4)\ W(4,1) \\
& = {\rm i} \times {\rm i}
\int {\rm d}(34)\ 
G(13)\ W(23)\ G(34)\ G(42)\ W(41).
\end{split}
\end{equation}
The different components $\Sigma^{\eta_1 \eta_2}$ of the self-energy 
on the time-loop contour (with $\eta_{1,2}=\pm$) are then given by
\begin{equation}
\begin{split}
\Sigma^{\eta_1 \eta_1}(12) = -\sum_{\eta_3 \eta_4} \ & \eta_3 \eta_4
\int {\rm d}(34)\  G^{\eta_1 \eta_3}(13)\ W^{\eta_2 \eta_3}(23) \\ 
& G^{\eta_3 \eta_4}(34)\ G^{\eta_4 \eta_2}(42)\ W^{\eta_4 \eta_1}(41).
\end{split}
\end{equation}
This self-energy corresponds to the so-called double-exchange
diagram. Note that we have studied the effects of such a diagram in
the different context of a propagating electron coupled to a local
vibration mode, in which the bosonic propagator $W$ is replaced by a
phonon propagator $D$\cite{Dash:2010}.

At the $\Gamma_{(1)}$ level of approximation, we find that the
polarizability is given by
\begin{equation}
\begin{split}
\tilde{P}(12) & = -{\rm i} \int {\rm d}(34)\
G(13)\  G(41)\ \Gamma_{(1)}(34;2) \\
& = 
\int {\rm d}(34)\
G(13)\  G(41)\ W(43)\ G(24)\ G(32) ,
\end{split}
\end{equation}
with components on $C_K$ given by
\begin{equation}
\begin{split}
\tilde{P}^{\eta_1 \eta_2}(12)
= \sum_{\eta_3 \eta_4} \ & \eta_3 \eta_4
\int {\rm d}(34) 
G^{\eta_1 \eta_3}(13)\  G^{\eta_4 \eta_1}(41) \\
& W^{\eta_4 \eta_3}(43)\ 
G^{\eta_2 \eta_4}(24)\ G^{\eta_3 \eta_2}(32). 
\end{split}
\end{equation}
Here again, and as well as for the self-energy, the retarded (advanced) part 
$\tilde{P}^r(12)$ is obtained from 
$\tilde{P}^r=\tilde{P}^{++}-\tilde{P}^{+-}$.
One can then express $\tilde{P}^r$ and $\tilde{P}^{+-}$ in a more compact form 
involving only terms like $X^{r,a,\lessgtr}$ (with $X \equiv G, W$).

\section{Time-ordered Green's functions at equilibrium}
\label{app:timeorderedGFs}

In this section we discuss in detail the relation between 
time-ordered Green's function (in energy representation) 
for two temperature limits. Differences are expected to arise
as shown in Chapter IV.17. of Ref.~[\onlinecite{Hedin:1969}].
We use the conventional equilibrium many-body perturbation theory (MBPT) 
to determine the time-ordered Green's function $G^t$, and the generalization 
of the Green's function onto the Keldysh time-loop contour at equilibrium
to determine the counterpart of the time-ordered Green's function $G^{++}$.

From MBPT, the time-ordered Green's function satisfies the Dyson-like
equation $G^t=g^t + g^t \Sigma^t G^t$ and the corresponding
time-ordered Green's function obtained from the Keldysh time-loop
expansion satisfies the corresponding Dyson-like equation $G^{++} =
g^{++} + (g \Sigma G)^{++}$.  In principle, from the conventional
definition we have $g^t = g^{++}$ and and should have $G^t = G^{++}$.

It is easy to show that from the rules of analytical continuation 
$G^{++} = g^{++} + (g \Sigma G)^{++}$
is expanded as follows
\begin{equation}
\label{eq:app_G++}
\begin{split}
G^{++} & = g^{++} 
	 + g^{++} \Sigma^{++} G^{++} \\
       & - g^{++} \Sigma^< G^>
	 + g^< \Sigma^> G^{++}
	 + g^< \Sigma^{--} G^{++} ,
\end{split}
\end{equation}
and after further manipulation (using the notation $(g/G)^t = (g/G)^{++}$),
\begin{equation}
\label{eq:app_G++bis}
G^t 	= g^t 
	+ \left( g^t \Sigma^t - g^< \Sigma^> \right) G^t
	- (g \Sigma)^< G^> .
\end{equation}
So, strictly speaking, the non-equilibrium formalism introduces two
extra terms $g^< \Sigma^> G^t$ and $(g \Sigma)^< G^>$ in the Dyson
equation for $G^t$. 

We now analyze these two terms in more detail.  First of
all, we recall that at equilibrium or in a steady state, the Green's
functions and self-energies depend only on the time difference of
their argument and can be Fourier transformed with a single energy
argument.  We then have the following expression 
\begin{equation}
\label{eq:app_G++ter}
\begin{split}
G^t(\omega) =
g^t(\omega) & + \left( g^t(\omega) \Sigma^t(\omega) - g^<(\omega)
  \Sigma^>(\omega) \right) G^t(\omega) \\
& - (g\Sigma)^<(\omega)
G^>(\omega) .
\end{split}
\end{equation}

Furthermore, at equilibrium or in a steady state, the lesser and
greater components of either a Green's function or a self-energy ($X^\lessgtr$)
can be expressed in terms of the corresponding advanced and retarded
quantity and a distribution 
function \cite{Ness:2010,Kita:2010,Meden:1995,Lipavski:1986}, i.e.
\begin{equation}
\label{eq:app_NEdistrib}
X^\lessgtr(\omega) =
-f^\lessgtr(\omega) (X^r(\omega)-X^a(\omega)) .
\end{equation}

At equilibrium $f^\lessgtr(\omega)=f_0^\lessgtr(\omega)$ and for a
system of fermoins, $f_0^<$ is given by the Fermi-Dirac distribution
function $f^{\rm eq}(\omega)=1/(1+\exp\beta(\omega-\mu^{\rm eq}))$
and $f_0^>=f^{\rm eq}-1$ (with $\beta=1/kT$).

At zero temperature, the Fermi-Dirac distribution takes only two
different values, $f^{\rm eq}=1$ or $0$. Hence we have the
property $(f^{\rm eq})^2=f^{\rm eq}$, which implies that
$f_0^<(\omega)f_0^>(\omega)=f^{\rm eq}(f^{\rm eq}-1)=0$.  Consequently
any products of the kind $X^<(\omega) Y^>(\omega)$ or $X^>(\omega)
Y^<(\omega)$ vanish.  Therefore we recover from the Keldysh time-loop
formalism Eq.~(\ref{eq:app_G++ter}) at zero temperature, 
the conventional Dyson equation $G^t=g^t + g^t\ \Sigma^t\ G^t$
as expected.

At finite temperature $f_0^<(\omega)f_0^>(\omega)=f^{\rm
  eq}(f^{\rm eq}-1)=kT \partial_\omega f^{\rm eq}\ne 0$, and the
product $f_0^< f_0^>$ gives a sharply peaked function at the Fermi level
$\mu^{\rm eq}=\varepsilon_F$ with a width of approximately $kT$.

We now check the individual contribution of each term $g^< \Sigma^>$
and $(g \Sigma)^< G^>$, first for a specific case (i.e. the quasi-particle
approximation) and then for the general case. 

In a quasi-particle scheme, i.e.\ when a single
index $k$ is good enough to represent the quantum states (with
energy $\varepsilon_k$), the Green's functions and the self-energies in
the absence and in the presence of interaction are diagonal in this
representation.  We have
\begin{equation}
\label{eq:app_1srcorterm}
\begin{split}
& g^<_k(\omega) \Sigma^>_k(\omega) \\
& = -f_{0,k}^<(g^r_k-g^a_k)(\omega) \times -f_{0,k}^>(\Sigma^r_k-\Sigma^a)(\omega) \\
& = 4\pi f_{0,k}^< f_{0,k}^> \delta( \omega - \varepsilon_k)\ \Im m \Sigma^r_k(\omega) .
\end{split}
\end{equation}

For purely fermionic systems at equilibrium, one usually has $\Im m
\Sigma^r_k(\mu^{\rm eq})=0$ \cite{Benedict:2002}, and therefore
$g^<_k(\mu^{\rm eq}) \Sigma^>_k(\mu^{\rm eq})=0$.  
When $\Im m \Sigma^r_k$ also vanishes in
the energy window around the Fermi level, defined by $f_0^< f_0^> \ne
0$, then the product $g^<_k(\omega) \Sigma^>_k(\omega)$ also vanishes. 
When there are no eigenvalues $\varepsilon_k$ (of the non-interacting system)
within this energy window, then once more we have $g^<_k(\omega) \Sigma^>_k(\omega)\sim 0$. 

Otherwise
$g^<_k(\omega) \Sigma^>_k(\omega) = \tilde Z_k \delta( \omega -
\varepsilon_k)$ with $\tilde Z_k = 4\pi \left( f_0^< f_0^> \Im m
  \Sigma^r_k(\omega)\right)_{\omega=\varepsilon_k}$.

For the second correction term, we have
\begin{equation}
\label{eq:app_2ndcorterm}
\begin{split}
& (g \Sigma)_k(\omega)^< G^>_k(\omega) \\
& = -f_{0,k}^<((g\Sigma)^r_k - (g\Sigma)^a_k)(\omega) \times -f_{0,k}^>(G^r_k-G^a_k)(\omega) \\
& = f_{0,k}^<(g^r_k \Sigma^r_k - g^a_k \Sigma^a_k)(\omega)\ f_{0,k}^>(G^r_k-G^a_k)(\omega) .
\end{split}
\end{equation}

For the quasi-particle scheme, $\Im m \Sigma^{r/a}_k \sim \pm {\rm
  i} \eta$ around the Fermi level $\mu^{\rm eq}\pm kT$, and we find that
\begin{equation}
\label{eq:app_2ndcorterm_bis}
(g_k \Sigma_k)^< G^>_k = -4\pi f_0^< f_0^> Z_k \Re e\Sigma^r_k(\varepsilon_k)\
\delta( \omega - \varepsilon_k) \delta( \omega - \tilde\varepsilon_k)
\end{equation}
with $Z_k^{-1}=1-(\partial \Re e \Sigma^r_k/\partial \omega)_{\omega=\tilde\varepsilon_k}$ 
being the effective mass renormalisation parameter and
$\tilde\varepsilon_k=\varepsilon_k+\Re e \Sigma^r_k$ being the renormalized eigenvalue.  
Hence the product $(g_k \Sigma_k)^< G^>_k$ vanishes because in general one has
$\tilde\varepsilon_k \ne \varepsilon_k$. In the opposite case when 
$\tilde\varepsilon_k = \varepsilon_k$ for some quantum states, the product $(g_k \Sigma_k)^< G^>_k$ 
also vanishes because then $\Re e \Sigma^r_k=0$.

Therefore our analysis show that, in the quasi-particle scheme at finite temperature, 
Eq.~(\ref{eq:app_G++ter}) reduces to the conventional Dyson equation
$G^t_k=g^t_k + g^t_k\ \Sigma^t_k\ G^t_k$ as expected.

Now we need to check what is happening to the two contributions $g^< \Sigma^>$
and $(g \Sigma)^< G^>$ beyond the quasi-particle approximation.
For that we can proceed further: going back to the full time-dependence of Eq.~(\ref{eq:app_G++bis})
and factorizing the non-interacting time-ordered Green's function $g^t$:
\begin{equation}
\label{eq:app_Gt_last}
G^t 	= g^t \left( 1 + \left( \Sigma^t - (g^t)^{-1} g^< \Sigma^> \right) G^t
	- (g^t)^{-1} (g \Sigma)^< G^> \right),
\end{equation}
with $(g \Sigma)^< = g^< \Sigma^a + g^r \Sigma^<$.

By using the equation of motion of the non-interacting time-ordered
Green's function $g^t$:
\begin{equation}
\label{eq:app_EOM_gt}
\left({\rm i} \frac{\partial}{\partial t_1} - h_0(1)\right) g^t(12) = \delta(12) ,
\end{equation}
it is straightforward to find that
\begin{equation}
  \label{eq:app_inv_gt}
(g^t)^{-1}(13)=\left({\rm i} \frac{\partial}{\partial t_1} - h_0(1)\right) \delta(13) .
\end{equation}
and consequently 
\begin{equation}
  \label{eq:app_inv_gt_time_gless}
\begin{split}
(g^t)^{-1} g^<(14) & \equiv \int {\rm d}3\ (g^t)^{-1}(13)
g^<(34) \\
& = \left({\rm i} \frac{\partial}{\partial t_1} - h_0(1)\right)
g^<(14) = 0, 
\end{split}
\end{equation}
the last equality comes from the definition of
$g^<(14)$.  Similarly one can find that $(g^t)^{-1} g^r \equiv \int
{\rm d}3\ (g^t)^{-1}(13) g^r(34)= \delta(14)$.

Hence Eq.~(\ref{eq:app_Gt_last}) is transformed into
\begin{equation}
\label{eq:app_Gt_last_bis}
G^t 	= g^t + g^t \Sigma^t G^t - \Sigma^< G^> ,
\end{equation}
where the last term $\Sigma^< G^>$ satisfies the detailed balance
equation at equilibrium \cite{Danielewicz:1984}: $\Sigma^< G^> =
\Sigma^> G^<$.  

Eq.~(\ref{eq:app_Gt_last_bis}) is the most general expression for $G^t$
and is the most important result of this section.
It is interesting to note that Eq.~(\ref{eq:app_Gt_last_bis}) is the equivalent 
of Eq.~(17.9) derived in Ref.~[\onlinecite{Hedin:1969}].  
However in our approach, the extra term $\Sigma^< G^>$ is obtained rigorously
from the use of the general Keldysh time-loop contour formalism. While in 
Ref.~[\onlinecite{Hedin:1969}], Hedin and Lundqvist introduced this 
correction term {\it ad hoc} in the Dyson equation for the finite temperature time-ordered 
Green's function in order to
recover the proper limit of the independent particle case.

Once more one can show that, after Fourier transforming,
the product $\Sigma^< G^>$ vanishes at equilibrium and at zero temperature
because of  Eq.~(\ref{eq:app_G++ter}) and $f_0^< f_0^>=0$.
Within the quasi-particle scheme at finite temperature, 
we have $\Sigma^<_k(\omega) G^>_k(\omega) = -4
f_{0,k}^< f_{0,k}^>\ \Im m \Sigma^r_k(\omega) \ \Im m G^r_k(\omega)$.
Thus, one needs to check the contributions of the spectral information in
$\Im m \Sigma^r_k(\omega)$ and in $\Im m G^r_k(\omega)$ (in the energy 
window defined by $f_{0,k}^< f_{0,k}^>$ around the Fermi level) to see
if the product $\Sigma^<_k G^>_k$ vanishes (as shown above).

We conclude this appendix by saying that there is indeed a difference
between the Dyson equations for the time-ordered Green's functions at
zero and finite temperature \cite{Hedin:1969,Fetter:1971}$^,$\footnote{The conventional equilibrium Green's 
formalism at finite temperature contains terms that are never considered at zero temperature, 
see page 289 of Ref.~[\onlinecite{Fetter:1971}].}. 
This result by no means contradicts the fact
that the Green's functions on the Keldysh contour, the time-ordered Green's function at zero temperature and 
the Matsubara temperature Green's function of imaginary argument all obey the same formal Dyson equation. 
Our derivations provide a rigorous mathematical result for the finite temperature time-ordered Green's 
function (in the energy representation)  which satisfies a Dyson equation with an extra term 
as introduced in an {\it ad-hoc} way in Chap IV.17. of Ref.~[\onlinecite{Hedin:1969}].

In our calculations, the correction term $\Sigma^< G^>$ is automatically
taken into account since we work with the Keldysh time-loop formalism.
We have checked numerically that the $\Sigma^< G^>$ indeed vanishes at zero
temperature. For finite temperatures we have found that $\Sigma^< G^> \sim 0$
in the energy window defined by $f_0^< f_0^> \ne 0$ since most of the
spectral weight is far below the Fermi level (see Figures \ref{fig:Aw_rs2} to \ref{fig:Aw_rs4_Tdep_lesspts}). 
However, in the limit of very high temperatures (i.e. $\omega_p/kT \ll 1$), the energy window defined by 
$f_0^< f_0^> \ne 0$ is wide and the product $\Sigma^< G^>$ does not vanish;
though the corrections are two orders of magnitude smaller than the amplitude
of the Green's function $G^t$ itself.

It would be interesting to find real cases of interacting electron
systems (probably of low dimensionality) for which the correction
term $\Sigma^< G^>$ is not negligible. At finite but low temperatures,
systems with a strong spectral density around the Fermi level
(i.e. presenting the Kondo effect) at low temperature should be a good
example. The high temperature limit for metallic systems represents another
interesting case as shown, for example, in Ref.~[\onlinecite{Benedict:2002}].

\end{document}